\documentclass[11pt,a4paper]{article}
\pdfoutput=1

\usepackage{jheppub}
\renewcommand\beforetochook[1]{}
\renewcommand\afterTocSpace[1]{}
\renewcommand\afterTocRuleSpace{}
\usepackage{bbm,bbold}
\usepackage{graphicx}

\usepackage{titlesec}
\titleformat{\paragraph}[hang] {\normalfont\bfseries } {} {0pt} {\boldmath}
\titlespacing*{\paragraph}{0pt}{8pt}{4pt}

\DeclareMathOperator\re{\text{Re}}
\DeclareMathOperator\Tr{\text{tr}}
\DeclareMathOperator\tr{\text{tr}}
\DeclareMathOperator\diag{\text{diag}}

\renewcommand\phi\varphi

\newcommand{\dphi}{\phi^\dagger}
\newcommand{\BZ}{_\text{BZ}}
\newcommand\vp{\mathbf{p}}
\newcommand{\SU}{\text{SU}}
\newcommand{\U}{\text{U}}
\newcommand{\mH}{\mathcal{H}}
\newcommand\1{\mathbb{1}}
\newcommand\Nb{N_b}
\newcommand\Nc{N_c}

\newcommand\Ngen{n}
\newcommand\mb{m}
\newcommand{\gi}{g_{I}}
\newcommand{\tgi}{\tilde g_I}
\newcommand{\gw}{g_W}

\title{Induced QCD I: Theory}

\author[a,b]{Bastian B. Brandt,}
\author[b]{Robert Lohmayer,}
\author[b]{and Tilo Wettig}
\affiliation[a]{Institute for Theoretical Physics, Goethe-University of Frankfurt,
  60438 Frankfurt, Germany}
\affiliation[b]{Institute for Theoretical Physics, University of Regensburg,
  93040 Regensburg, Germany}
\emailAdd{bastian.brandt@ur.de}
\emailAdd{robert.lohmayer@ur.de}
\emailAdd{tilo.wettig@ur.de}

\abstract{We explore an alternative discretization of continuum
  $\SU(\Nc)$ Yang-Mills theory on a Euclidean spacetime lattice,
  originally introduced by Budzcies and Zirnbauer. In this
  discretization the self-interactions of the gauge field are induced
  by a path integral over $\Nb$ auxiliary boson fields, which are
  coupled linearly to the gauge field. The main progress compared to
  earlier approaches is that $\Nb$ can be as small as $\Nc$.  In the
  present paper we (i) extend the proof that the continuum limit of
  the new discretization reproduces Yang-Mills theory in two
  dimensions from gauge group $\U(\Nc)$ to $\SU(\Nc)$, (ii) derive
  refined bounds on $\Nb$ for non-integer values, and (iii) perform a
  perturbative calculation to match the bare parameter of the induced
  gauge theory to the standard lattice coupling.  In follow-up papers
  we will present numerical evidence in support of the conjecture that
  the induced gauge theory reproduces Yang-Mills theory also in three
  and four dimensions, and explore the possibility to integrate out
  the gauge fields to arrive at a dual formulation of lattice QCD.
}

\begin{document}

\maketitle

\section{Introduction}

Strong-coupling approaches to lattice gauge theories, in particular to
lattice QCD, have a long history since they allow both for analytical
investigations and for the construction of new simulation algorithms,
see, e.g.,
\cite{Blairon:1980pk,KlubergStern:1981wz,Kawamoto:1981hw,Rossi:1984cv,Karsch:1988zx}.
Typically, these approaches work only if the self-interaction of the
gauge fields is neglected, giving rise to an uncontrolled systematic
error of the results.  There have been several ideas
\cite{Bander:1983mg,Hamber:1983nm,Kazakov:1992ym,Hasenfratz:1992jv} to
overcome this limitation by minimally coupling auxiliary degrees of
freedom to the gauge field such that, after the auxiliary fields are
integrated out, the correct gauge action is ``induced'' in a
well-defined limit.  However, in most cases this limit involves an
infinite number of auxiliary fields, which is not useful for
numerical simulations.  An exception is the approach of Kazakov and
Migdal (KM) \cite{Kazakov:1992ym}, but the KM model does not induce
the desired Yang-Mills (YM) theory since (i) the action has a local
center symmetry, which forces all Wilson loops to vanish
\cite{Kogan:1992sn,Elitzur:1975im}, and (ii) an explicit solution with
a quadratic potential showed that in this case the KM model does not
yield the correct continuum behavior \cite{Gross:1992dy}, see
\cite{Weiss:1994gn} for a review.

A major step forward was taken more than 10 years ago by Budczies and
Zirnbauer (BZ) \cite{Budczies:2003za}, who presented a novel method to
induce the gauge action.  The essential ideas of this method, which
uses a small number $\Nb$ of auxiliary bosons, will be reviewed in
section~\ref{sec:BZ}.  In short, the BZ method uses a ``designer action''
that couples the auxiliary bosons to the gauge field in such a way
that, if the boson mass is taken to a critical value, the theory has
the same continuum limit as YM theory provided that $\Nb$ is larger
than a certain lower bound.  This was shown analytically for $d=2$ and
gauge group $\U(\Nc)$ by matching to an earlier result of Witten
\cite{Witten:1991we}, while for $d>2$ there is no analytical proof but
a plausible universality argument.

In this paper we study various theoretical aspects of the BZ method.
In section~\ref{sec:induced} we reformulate the BZ action to eliminate a
spurious sign problem.  In section~\ref{sec:limit} we generalize it to
the case of gauge group $\SU(\Nc)$ and derive, for both $\U(\Nc)$ and
$\SU(\Nc)$, precise bounds on $\Nb$ for which a continuum limit exists
and for which this limit corresponds to YM theory.  In
section~\ref{sec:pert} we perform a perturbative calculation for
$\SU(\Nc)$ to match the coupling in the BZ action to that of the
standard Wilson plaquette action.

In follow-up papers we will present numerical evidence for the
conjecture that the BZ method induces YM theory in the continuum limit
by means of lattice simulations for $\SU(2)$ in 3d and for $\SU(3)$ in
4d, and explore the construction of dual formulations of lattice QCD
by applying the BZ method and integrating out the gauge and fermion
fields.  First reports of our study appeared in~\cite{Brandt:2014rca,Brandt:2015nql}.

\section{Boson-induced pure gauge theory}
\label{sec:induced}

In this section we review the basic idea of the BZ method and
reformulate the action to eliminate a spurious sign problem.  Unless
stated otherwise, all dimensionful quantities are made dimensionless
by multiplication with an appropriate power of the lattice spacing
$a$, which we set to unity.

\subsection{Formulation of Budczies and Zirnbauer}
\label{sec:BZ}

\begin{figure}
  \centering
  \includegraphics{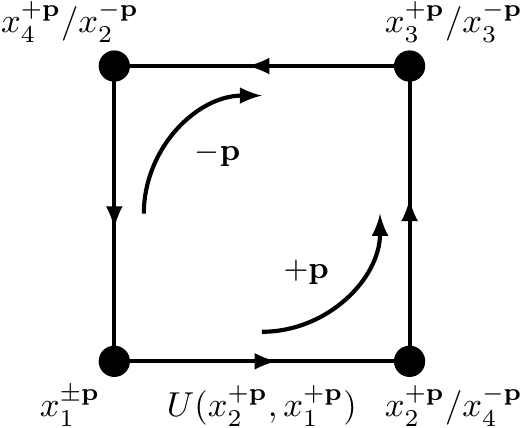}
  \caption{Conventions for the orientations of a plaquette and the
    corresponding lattice sites and link variables.  With
    $U_\mu(x)=U(x+\mu,x)$ we have $U_{\vp} = U(x^\vp_1,x^\vp_4)
    U(x^\vp_4,x^\vp_3) U(x^\vp_3,x^\vp_2) U(x^\vp_2,x^\vp_1)$ and a
    similar expression for $U_{-\vp}$ with $\vp$ replaced by $-\vp$.
    Because of $U(x+\mu,x) = U(x,x+\mu)^\dagger$ we have
    $U_{-\vp}=U_{\vp}^\dagger$.}
  \label{fig:plaquettes}
\end{figure}

We restrict ourselves to gauge group $G=\U(\Nc)$ or $\SU(\Nc)$ in this
paper.  In \cite{Budczies:2003za} the gluonic weight for a
configuration $[U]$ of gauge fields $U_\mu(x)$ in the fundamental
representation of $G$ is taken to be
\begin{align}
  \label{eq:BZ-det-action}
  \omega\BZ[U] \sim \prod_{\pm\vp} \det\big(m\BZ^4-U_{\vp}\big)^{-\Nb}
  =\prod_{p} \big| \det\big(m\BZ^4-U_p\big) \big|^{-2\Nb} \,,
\end{align}
where here and below the symbol $\sim$ means that we have ignored a
normalization factor that will be reinstated when it becomes
important.  The first product is over all oriented plaquettes (see
figure~\ref{fig:plaquettes} for our conventions),\footnote{For the sake
  of brevity and clarity we restrict ourselves to a hypercubic
  lattice.  The discussion can straightforwardly be generalized to
  other lattice geometries using the framework and notation
  of~\cite{Budczies:2003za}.}
while the second product is over all unoriented plaquettes, where we have
used $U_{-\vp}=U_{\vp}^\dagger$ and defined $U_p=U_\vp$.  (The
definition $U_p=U_{-\vp}$ would lead to the same final results.)  As a
special case of the more general discussion in~\cite{Budczies:2003za},
we take $m\BZ$ to be real and identical for all plaquettes.
Equation~\eqref{eq:BZ-det-action} implies that we need $m\BZ>1$ to
have a convergent theory.

Note that the weight factor~\eqref{eq:BZ-det-action} has the same
global center-symmetry property as in YM theory due to the fact that
the gauge fields only appear as full plaquettes in a class function in
the action. This is of particular relevance for the deconfinement
transition. The invariance under a local center symmetry is a major
shortcoming of the KM model.

The expectation value of an observable $O[U]$ is given by the path
integral
\begin{align}
  \label{eq:pint}
  \left\langle O \right\rangle = \frac{1}{Z} \int_{G} [dU] \, O[U]
  \omega\BZ[U]
\end{align}
with the partition function $Z$ defined by $\langle 1 \rangle =1$.  We
follow the convention of \cite{Budczies:2003za} to use square brackets
for a full set (i.e., a configuration) of gauge (or auxiliary) fields.

There is a ``naive'' pure-gauge limit in which the effective action
resulting from \eqref{eq:BZ-det-action} reduces to the Wilson plaquette
action~\cite{Wilson:1974sk}.  Writing
$\omega\BZ[U] \sim \exp\left(-S\BZ^{\text{eff}}[U]\right)$ we have
\cite{Budczies:2003za}
\begin{align}\label{eq:BZ}
  S\BZ^{\text{eff}}[U] \to S_W[U] = -\frac{\beta}{\Nc} \sum_p
  \re\tr U_p
\end{align}
in the combined limit $m\BZ\to\infty$ and $\Nb\to\infty$ with
$\beta=2\Nc\Nb/m\BZ^4$ fixed.  The continuum limit then corresponds to
taking the lattice coupling parameter $\beta$ to infinity.  The naive
limit requires large $\Nb$, similar as in the earlier approaches
\cite{Bander:1983mg,Hamber:1983nm,Hasenfratz:1992jv}.  However, the
main point of the BZ method is that YM theory can also be obtained in
the ``interesting'' limit $m\BZ\to1$ with $\Nb$ fixed at a finite (and
small) value.  This will be the subject of section~\ref{sec:limit}.

To bosonize the determinants in \eqref{eq:BZ-det-action} we assume
that $\Nb$ is a positive integer and introduce auxiliary boson fields
$\phi_{b,\vp}(x_j^\vp)$, where the index $b=1,\ldots,\Nb$ labels the
boson flavor, $x_j^\vp$ with $j=1,\ldots,4$ is defined in
figure~\ref{fig:plaquettes}, and the index $\vp$ on $\phi$ means that
we have different fields for different (oriented) plaquettes. The
fields are in the fundamental representation of $G$ and thus carry a
color index that we did not write explicitly.  Using these fields we
have
\begin{align}
  \label{eq:BZ-action-weight}
  \omega\BZ[U] \sim \int [d\phi] \,
  \exp\left(-S\BZ[\phi,U]\right)
\end{align}
with
\begin{align}
  \label{eq:BZ-action}
  S\BZ [\phi,U] = \sum_{b=1}^{\Nb} \sum_{\pm \vp} \sum_{j=1}^{4}
  \big[ m\BZ \dphi_{b,\vp}(x_j^\vp)\phi_{b,\vp}(x_j^\vp) -
  \dphi_{b,\vp}(x_{j+1}^\vp) U(x_{j+1}^\vp,x_j^\vp)
  \phi_{b,\vp}(x_j^\vp) \big] \,,
\end{align}
where $x_5^\vp\equiv x_1^\vp$.  The connection between
\eqref{eq:BZ-det-action} and~\eqref{eq:BZ-action-weight} is readily
shown by rewriting the action~\eqref{eq:BZ-action} in bilinear form,
integrating out the boson fields, and converting the matrix in the
resulting determinant to upper-triangular form.  From
\eqref{eq:BZ-action} it is clear that the parameter $m\BZ$ is the mass
of the auxiliary bosons and that the total number of boson fields per
plaquette is $2\Nb$.

\subsection{Alternative formulation without sign problem}

One of the interesting aspects of the BZ method is that it can lead to
reformulations of lattice QCD.  To make numerical
simulations feasible it is important to have a real action so that the
weight factor is real and positive.  While the weight
factor~\eqref{eq:BZ-det-action}, in which the bosons are integrated
out, satisfies this criterion, this is no longer true for the
action~\eqref{eq:BZ-action}.  In this case the action is generically
complex since the imaginary parts of the terms containing the
positively and negatively oriented links only cancel after averaging
over the boson fields.

The sign problem in the action~\eqref{eq:BZ-action} would not be
present if the two hopping terms including a particular link
$U_\mu(x)$, i.e., the terms
\begin{align}
  \label{eq:crit-terms}
  \dphi_{b,+\vp}(x+\mu) U_\mu(x) \phi_{b,+\vp}(x) \quad \text{and}
  \quad \dphi_{b,-\vp}(x) U_\mu(x)^\dagger \phi_{b,-\vp}(x+\mu)\,,
\end{align}
were complex conjugates of each other so that their sum is real.  This
can be achieved by assigning the boson fields to unoriented (rather
than oriented) plaquettes and using the alternative action
\begin{align}
  \label{eq:action}
  S_B [\phi,U] = \sum_{b=1}^{\Nb} \sum_{p} \sum_{j=1}^{4}
  \Big[&\mb\dphi_{b,p}(x^p_j)\phi_{b,p}(x^p_j) -
  \dphi_{b,p}(x^p_{j+1})
  U(x^p_{j+1},x^p_j)\phi_{b,p}(x^p_j) \Big. \notag\\
  & \Big. - \dphi_{b,p}(x^p_{j}) U(x^p_j,x^p_{j+1})
  \phi_{b,p}(x^p_{j+1})\Big]
\end{align}
with $x^p_j=x^{+\vp}_j$.  Note that we now have only half the number
of boson fields compared to the action~\eqref{eq:BZ-action}. The
matrix associated with the bilinear form in the boson fields is
Hermitian, leading to a real action and thus a positive definite
weight
\begin{align}
  \label{eq:action-weight}
  \omega[\phi,U] \sim \exp\left(-S_B[\phi,U]\right)
\end{align}
for all gauge and boson field configurations.

Some algebra is needed to show that the two
actions~\eqref{eq:BZ-action} and~\eqref{eq:action} are equivalent.
The first step is to integrate out the boson fields in the path
integral associated with the action~\eqref{eq:action}. This yields the
inverse determinant of the matrix corresponding to the bilinear form
in the boson fields. This matrix is diagonal in $b$ and $p$ so that
its determinant factorizes into a product of determinants of
$4\Nc\times 4\Nc$ matrices, with the product running over plaquettes
and boson flavors.  These determinants can be evaluated by converting
the $4\times4$ part with $\Nc\times \Nc$ matrices as entries to upper
triangular form. The final result is
\begin{align}
  \label{eq:det-action}
  \omega[U] \sim \prod_{p}
  \det\Big(1-\frac\alpha2\big(U_p+U_p^\dagger\big)\Big)^{-\Nb}
\end{align}
with\footnote{Note that the result~\eqref{eq:mtilde} for $\alpha$ is
  only valid for hypercubic lattices. The relation between $\alpha$
  and $\mb$ will be different for other lattice structures.}
\begin{align}
  \label{eq:mtilde}
  \frac2\alpha = \mb^4 - 4\mb^2 + 2 \,.
\end{align}
The weight factors~\eqref{eq:BZ-det-action} and~\eqref{eq:det-action}
are directly related via
\begin{align}
  \label{eq:equivalence}
  \big| \det\big(m\BZ^4-U_p\big) \big|^{2} & =
  \det\Big(\big(m\BZ^4-U_p\big)\big(m\BZ^4-U_p^\dagger\big)
  \Big) \notag\\
  & \sim \det\left(\frac2\alpha-\big(U_p+U_p^\dagger\big)\right)
\end{align}
if we identify
\begin{align}
  \label{eq:mtwid}
  \frac2\alpha = m\BZ^4 + m\BZ^{-4} \,.
\end{align}
From \eqref{eq:mtwid} and $m\BZ>1$ we obtain the condition
$0<\alpha<1$. Via \eqref{eq:mtilde} this implies $m>2$ in
\eqref{eq:action}. The limit $m\BZ\to 1^+$ corresponds to the limit
$\alpha\to 1^-$ (or, equivalently, $m\to2^+$).  From now on we will
use the weight factor~\eqref{eq:det-action}.

\section{\boldmath Continuum limit for $\U(\Nc)$ and $\SU(\Nc)$}
\label{sec:limit}

In \cite{Budczies:2003za} it was shown, for $G=\U(\Nc)$, that a
continuum limit exists for $\alpha\to1$ provided that $\Nb\ge\Nc$, and
that in two dimensions this continuum limit is in the YM universality
class if $\Nb\ge\Nc+1$.  For three or more dimensions it was
conjectured, based on plausible universality arguments, that the
continuum limit is in the YM universality class if $\Nb\ge\Nc$.

In this section we do not assume that $\Nb$ is integer and also
consider $G=\SU(\Nc)$.  In section~\ref{sec:charexp} we set up the basic
mathematical framework.  In section~\ref{sec:delta} we derive refined
bounds on $\Nb$ for which the gauge theory with weight
function~\eqref{eq:det-action} has a continuum limit for $\alpha\to1$.
In section~\ref{sec:equiv} we derive bounds on $\Nb$ which ensure that
the continuum limit is in the YM universality class.  The main results
are given in \eqref{eq:delta} and~\eqref{eq:cont}.

\subsection{Character expansion and exponential parameterization}
\label{sec:charexp}

Let us write the gluonic weight \eqref{eq:det-action} in the form
\begin{align}
  \label{eq:weight}
  \omega[U]\sim\prod_p\omega(U_p)\qquad\text{with}\qquad
  \omega(U)=\frac1{Z(\alpha)}\,{\det}\Big(1-\frac\alpha2\big(U+U^\dagger\big)\Big)^{-\Nb}\,.
\end{align}
For simplicity of notation we denote the weights for the ensemble of
gauge fields and for a single plaquette by the same symbol $\omega$.
The distinction between the two cases is made by the square or round
brackets.  The factor $Z(\alpha)$ ensures that $\omega(U)$ is properly
normalized and is therefore given by
\begin{align}
  \label{eq:Z}
  Z(\alpha)=\int_G dU\,
  {\det}\Big(1-\frac\alpha2\big(U+U^\dagger\big)\Big)^{-\Nb}\,,
\end{align}
where $dU$ denotes the Haar measure, normalized such that
$\int_GdU=1$.

Since $\omega(U)$ is a class function we can expand it in the
characters of the irreducible representations (or irreps) of $G$,
\begin{align}
  \label{eq:charexp}
  \omega(U)=\sum_\lambda c_\lambda(\alpha)\chi_\lambda(U)\,,
\end{align}
where $\lambda$ labels the inequivalent irreps and $\chi_\lambda(U)$
is the character of $U$ in $\lambda$.  Using character orthogonality,
\begin{align}\label{eq:char-orth}
  \int_G dU\,\chi_\lambda(U V) \chi_\gamma(U^\dagger W) = \delta_{\lambda
    \gamma} \frac{1}{d_\lambda} \chi_\lambda(VW)
\end{align}
with $d_\lambda=\dim(\lambda)$, 
the coefficients are given by
\begin{align}
  \label{eq:c}
  c_\lambda(\alpha)=\int_G dU\,\omega(U)\chi_\lambda(U^\dagger)\,.
\end{align}
Clearly $c_0(\alpha)=1$, where $\lambda=0$ labels the trivial
representation. When $\lambda$ is not specified, $\det(\cdot)$ and
$\tr(\cdot)$ always refer to the fundamental representation.

To compute the integral \eqref{eq:c} we use an exponential
parameterization of the form
\begin{align}
  \label{eq:gamma}
  U=\exp(i\sqrt \gamma H) \qquad\text{with }
  \gamma=\frac2\alpha(1-\alpha) \text{ and } H\in
  \mathfrak{g}\,,
\end{align}
where $\mathfrak g$ denotes the group algebra of $G$.  The factor of
$\sqrt \gamma$ was chosen such that the
parameterization~\eqref{eq:gamma} leads to a systematic expansion of
the $U$-dependent part of the weight function in powers of
$(1-\alpha)$ for fixed $H$, 
\begin{align}
  \label{eq:UH}
  \det\Big(1-\frac\alpha2\big(U+U^\dagger\big)\Big)
  &=\det\big(1-\alpha\cos(\sqrt\gamma H)\big)\notag\\
  &=(1-\alpha)^{\Nc}\det\Big(1+H^2-\frac\gamma{12}H^4+\ldots\Big)\,,
\end{align}
where the higher-order terms in the determinant are of the form
$\gamma^{k}H^{2k+2}$ with $k\geq 2$.

The integration measure becomes (see, e.g., \cite[App.~C]{rothe2005lattice})
\begin{align}\label{eq:dU}
  dU=\gamma^{\Ngen/2}\sqrt{\det g(H)}\,dH \quad\text{with}\quad
  g(H)=\sum_{\ell=0}^\infty\frac{(-1)^\ell\gamma^{\ell}}{(2\ell+2)!}\,\mH^{2\ell}\,,
\end{align}
where
\begin{align}
  \Ngen=
  \begin{cases}
    \Nc^2 & \text{for }G=\U(\Nc)\,,\\
    \Nc^2-1 & \text{for }G=\SU(\Nc)\\
  \end{cases}
\end{align}
is the number of generators of $G$,
$\mH=\sum_{a=1}^nh_at_a^{(\text{adj})}$ denotes the element of the
adjoint representation of $\mathfrak g$ corresponding to
$H=\sum_{a=1}^nh_at_a^{(\text{fund})}$ in the fundamental
representation, and the $t_a$ are the generators of the representation
normalized according to \eqref{Eq:GenNorm}.  The integral over $H$ is
defined as an $\Ngen$-dimensional integral over the coefficients $h_a$, i.e., 
\begin{align}
  \label{eq:dH}
  dH=\prod_{a=1}^ndh_a\,.
\end{align}
The integration domain $V(\gamma,\mathfrak g)$ is chosen such that the group
$G$ is covered exactly once (or a finite number of times, resulting in
an irrelevant normalization factor).  Note that this domain
$V(\gamma,\mathfrak g)$ depends on $\gamma$.  In fact, we will only
have to evaluate integrals of class functions, where the appropriate
integration domains for the eigenvalues of $U$ are obvious.

For the expansion of the character we have
\cite{Weyl:1997,Budczies:2003za}
\begin{align}\label{eq:chiU}
  \frac{\chi_\lambda(e^{-i\sqrt \gamma H})}{d_\lambda}=
  1-i\sqrt{\gamma}\frac{q(\lambda)}{\Nc}\tr H -\frac{\gamma}2\left(
    \left(\frac{q(\lambda)^2}{\Nc^2}-\frac{A(\lambda)}{\Nc}\right)(\tr
    H)^2 +A(\lambda)\tr H^2\right)+\ldots
\end{align}
with\footnote{Note that our $C_2^\U(\lambda)$ differs from
  \cite{Budczies:2003za} by a factor of $2$ since we use the standard
  normalization \eqref{Eq:GenNorm}.}
\begin{align}
  \label{eq:A}
  A(\lambda)&=\frac2{\Nc^2-1}\left(C_2^{\U}(\lambda)-\frac{q(\lambda)^2}{2\Nc}\right),\\
  q(\lambda)&=\sum_{j=1}^{\Nc}\lambda_j\,,\\
  C_2^{\U}(\lambda)&= \frac12
  \sum_{j=1}^{\Nc}\lambda_j(\lambda_j+\Nc+1-2j)\,.
\end{align}
Here, the ordered set of integers
$\lambda_1\ge\lambda_2\ge\ldots\ge\lambda_{\Nc}$ defines the irrep
$\lambda$ of $G$ \cite{Weyl:1997},\footnote{Two irreps $\lambda$ and
  $\mu$ related by $\lambda_j=\mu_j+m$ with $m\in\mathbbm Z$ only
  differ by a factor of $\det(U)^m$. For $G=\SU(\Nc)$ we have
  $\det U=1$. In this case $\lambda$ and $\mu$ are equivalent, and the
  inequivalent irreps are conventionally chosen to be those with
  $\lambda_{\Nc}=0$.  Then all inequivalent irreps $\lambda$ are given
  by Young diagrams with $\Nc-1$ rows, with $\lambda_j\ge0$ equal to
  the length of row $j$.}  and $C_2^\U(\lambda)$ is the quadratic
Casimir invariant of $\U(\Nc)$. The second factor on the RHS of
\eqref{eq:A} is the quadratic Casimir invariant of $\SU(\Nc)$
\cite{Perelomov1965},
\begin{align}
  C_2^{\SU}(\lambda)&= C_2^{\U}(\lambda)-\frac{q(\lambda)^2}{2 \Nc}\,.
\end{align}
For $G=\SU(\Nc)$ we have $\tr H=0$ so that \eqref{eq:chiU}
simplifies to
\begin{align}\label{eq:chiSU}
  \frac{\chi_\lambda(e^{-i\sqrt \gamma H})}{d_\lambda}= 1- \frac{
    C_2^{\SU}(\lambda)}{\Nc^2-1}\gamma\tr H^2+\ldots\,, \qquad
  G=\SU(\Nc)\,.
\end{align}
We now pull out a trivial factor of
$(\gamma/2)^{n/2}(1-\alpha)^{-\Nb\Nc}$ from both $Z(\alpha)$ in
\eqref{eq:Z} and $dU {\det}(1-\alpha \cos(\sqrt{\gamma} H))^{-\Nb}$ in
\eqref{eq:c} and obtain
\begin{align}
  \label{eq:c2}
  c_\lambda(\alpha)&=\frac{\bar c_\lambda(\gamma)}{\bar c_0(\gamma)}\,,\\
  \label{eq:cbar}
  \bar c_\lambda(\gamma)&=\int_{V(\gamma,\mathfrak g)}dH\,
  \bar\omega(H)\chi_\lambda(e^{-i\sqrt\gamma H})\,,\\
  \bar\omega(H)&=\sqrt{\det(2g(H))}
  \det\left(\frac{1-\alpha\cos(\sqrt\gamma
      H)}{1-\alpha}\right)^{-\Nb}\,,
  \label{eq:omegabar}
\end{align}
where $\alpha$ and $\gamma$ are related by \eqref{eq:gamma}.
Note that
\begin{subequations}\label{eq:limits}
  \begin{equation}
    \lim_{\alpha\to1}\sqrt{\det(2g(H))}=1\,,
    \label{eq:limit_g}
  \end{equation}
  \begin{equation}
    \lim_{\alpha\to1}\bar\omega(H)=\det(1+H^2)^{-\Nb}\,,
    \label{eq:limit_omega}
  \end{equation}
  \begin{equation}
    \lim_{\alpha\to1}\chi_\lambda(e^{-i\sqrt\gamma H})=\chi_\lambda(\1)
    =d_\lambda\,.
    \label{eq:limit_char}
  \end{equation}
\end{subequations}
                                                 
\subsection[$\delta$-function property]{\boldmath $\delta$-function
  property}
\label{sec:delta}

As explained in \cite{Budczies:2003za}, a continuum limit is obtained
if $\omega(U)$ approaches a $\delta$-function located at $U=\1$.  As a
consequence of the Peter-Weyl theorem, the character expansion of the
$\delta$-function is given by \eqref{eq:charexp} with
$c_\lambda(\alpha)$ replaced by the dimension $d_\lambda$ of the irrep
$\lambda$.  Hence, $\omega(U)$ becomes a $\delta$-function if
$\lim_{\alpha\to1}c_\lambda(\alpha)=d_\lambda$ for all $\lambda$.  We
now investigate under what conditions this criterion is satisfied for
the different gauge groups.

\subsubsection[$G=\U(1)$]{\boldmath $G=\U(1)$}\label{Sec:U1}

It is instructive to first study the simplest case $G=\U(1)$ in some
detail. In this case, \eqref{eq:cbar} reduces to the one-dimensional
integral
\begin{align}\label{eq:cbarU1}
  \bar c_\lambda(\gamma)= \int_{0}^{\pi /\sqrt\gamma} dx\,
  \frac1{(1+x^2)^{\Nb}} \sum_{k,m=0}^\infty a_{k,m}(\lambda)
  \gamma^{k+m} \left(\frac{x^4}{1+x^2}\right)^k x^{2m}
\end{align}
with some coefficients $a_{k,m}(\lambda)$.  The asymptotic
behavior of $\bar c_\lambda(\gamma)$ as $\alpha \to 1$, i.e.,
$\gamma\to 0$, is therefore determined by integrals of the form
\begin{align}\label{eq:Ikm}
  I_{k,m}(\gamma)= \gamma^{k+m} \int_0^{\pi/\sqrt\gamma} dx\,
  \frac{x^{2(k+m)+2k}}{(1+x^2)^{\Nb+k}} \,.
\end{align}
The integral in \eqref{eq:Ikm} is finite as $\gamma \to 0$ as long as
$\Nb>k+m+\frac12$. The limit is
\begin{align}\label{Eq:BinomialResult}
  \int_0^\infty dx\,
  \frac{x^{4k+2m}}{(1+x^2)^{\Nb+k}}=\frac{\Gamma\left(2k+m+\frac12\right)
    \Gamma\left(\Nb-k-m-\frac12\right)}{2\Gamma(\Nb+k)}\,.
\end{align}
For $\Nb<k+m+\frac12$, the integral in \eqref{eq:Ikm} diverges like
$\gamma^{-(k+m+\frac12-\Nb)}$. If $\Nb=k+m+\frac12$, we obtain a
logarithmic divergence. Hence, the leading-order behavior of
$I_{k,m}(\gamma)$ as $\gamma\to0$ is given by
\begin{align}
  \label{eq:Ikm_leading}
  I_{k,m}(\gamma) \propto
  \begin{cases}
    \gamma^{k+m}&\text{for } k+m < \Nb-\frac12\,,\\
    \gamma^{\Nb-\frac12} \log\gamma &\text{for } k+m= \Nb-\frac12\,,\\
    \gamma^{\Nb-\frac12}&\text{for } k+m > \Nb-\frac12\,.
  \end{cases}
\end{align}
From \eqref{eq:Ikm_leading} we see that for $\Nb > \frac12$, 
$\bar c_\lambda(\gamma)$ is dominated by the finite and non-zero term
with $k=m=0$ in \eqref{eq:cbarU1}, with corrections of order
$\mathcal O \big(\gamma^{\min(1,\Nb-\frac12)}\big)$, or
$\mathcal O \left(\gamma\log\gamma\right)$ for $\Nb=\frac32$.  For
$\Nb = \frac12$, the term with $k=m=0$ diverges like $\log\gamma$
while all other terms are finite as $\gamma\to 0$.  Therefore, for
$\Nb \geq \frac12$ we have
\begin{align}\label{eq:U1continuum}
  \lim_{\alpha\to1} c_\lambda(\alpha)= \lim_{\gamma\to 0} \frac{\bar
    c_\lambda(\gamma)}{\bar
    c_0(\gamma)}=\frac{a_{0,0}(\lambda)}{a_{0,0}(0)}=d_\lambda=1\,,
\end{align}
where the penultimate equality follows from \eqref{eq:limits}.  This
implies that for $\Nb\ge\frac12$ the weight function $\omega(U)$
reduces to a $\delta$-function on the $\U(1)$ manifold in the limit
$\alpha\to1$.

On the other hand, for $\Nb < \frac12$ we have
$k+m\geq 0 > \Nb -\frac12$ for all terms, i.e., for all $k$ and $m$ we
obtain the same leading divergence,
$I_{k,m}\propto \gamma^{-(\frac12-\Nb)}$. Therefore, all terms in the
sum over $k$ and $m$ contribute to
$\lim_{\alpha \to 1} c_\lambda(\alpha)$, making the dependence on
$\lambda$ non-trivial, and consequently
$\lim_{\alpha \to 1} c_\lambda(\alpha)\neq d_\lambda$ generically.
Hence we obtain a $\delta$-function if and only if
$\Nb\geq \frac12$.

For $\Nb\in\mathbbm N$, the coefficients $c_\lambda$ can be
calculated analytically as a function of $\alpha$, see
appendix~\ref{App:u1}. The results are consistent with the condition
$\Nb \geq \frac12$.

\subsubsection[$G=\SU(2)$]{\boldmath $G=\SU(2)$}

Parametrizing $H\in \mathfrak{su}(2)$ in terms of its eigenvalues,
$H=V \diag(x,-x) V^\dagger$ with $V\in \SU(2)$, leads to
$\det(1+H^2)=(1+x^2)^2$ and integration measure $dH\propto (x-(-x))^2$
(see section~\ref{sec:SUNdelta} below). The coefficient $\bar c_\lambda$
is then given by a single integral equivalent to the
integral~\eqref{eq:cbarU1} except for the replacement
$(1+x^2)^{-\Nb} \to x^2(1+x^2)^{-2\Nb}$.  We can therefore immediately
apply the power-counting arguments of the previous section after
substituting $\Nb \to 2 \Nb -1$. This results in $\Nb \geq \frac34$ as
a necessary and sufficient condition for $\omega(U)$ to approach the
$\delta$-function as $\alpha\to1$. 

For $2\Nb\in\mathbbm N$, the coefficients $c_\lambda$ can be
calculated analytically as a function of $\alpha$, see
appendix~\ref{App:su2}. The results are consistent with the condition
$\Nb \geq \frac34$.

\subsubsection[$G=\U(\Nc)$]{\boldmath $G=\U(\Nc)$}
\label{sec:UNdelta}

Since $H$ is Hermitian we transform to the eigenvalue representation
$H=V\diag(x_j)V^\dagger$ with $x_j$ ($1\leq j \leq \Nc$) real and $V$
unitary.  The Jacobian $J$ of this transformation is given by the square
of a Vandermonde determinant, $J=\prod_{j<k}(x_j-x_k)^2$.

Let us first determine the asymptotic behavior (for $\gamma\to0$) of
the integral of the determinant~\eqref{eq:limit_omega} over the domain
$V(\gamma,\mathfrak u(\Nc))$,
\begin{align}\label{eq:Uintegral}
  \int_{V(\gamma,\mathfrak{u}(\Nc))}dH\,\det(1+H^2)^{-\Nb} \propto
  \int_{-\pi/\sqrt\gamma}^{\pi/\sqrt\gamma}\Big(\prod_{j=1}^{\Nc}dx_j\Big)
  \Big(\prod_{j<k}(x_j-x_k)^2\Big)
  \Big(\prod_{j=1}^{\Nc}(1+x_j^2)^{-\Nb}\Big)\,.
\end{align}
We now split the integral over the eigenvalues $x_j$ into several
integrals over subdomains and separately analyze their asymptotic
behavior by simple power counting:

(i) In domains where all $x_j$ are ``finite'' (in the sense that they
are of order $\gamma^0$ and do not go to infinity like
$1/\sqrt\gamma$), the contributions to the integral are finite, i.e.,
of order $\gamma^0$.

(ii) Choose an integer $k$ with $0\leq k \leq \Nc-1$.  In domains
where $k$ of the variables (say, the $x_i$ with $1\leq i \leq k$) stay
``finite'' and the remaining $\Nc-k$ variables ($x_i$,
$k+1\leq i \leq \Nc$) are ``large'' (in the sense that they go to
infinity like $1/\sqrt\gamma$) and ``independent'' (in the sense that
generically differences $x_i-x_j$ for $k+1\leq i < j \leq \Nc$ are
``large''), the leading-order contributions to the integral are
proportional to
\begin{align}\label{eq:k-contr-U}
  \left( \frac 1{\sqrt\gamma}\right)^{\Nc-k+2\big[\binom\Nc2 -
  \binom k 2 \big]-2\Nb(\Nc-k)}
  =\gamma^{\frac12(k-\Nb)^2-\frac12(\Nc-\Nb)^2}\,,
\end{align}
provided the exponent on the RHS is negative (otherwise, the integral
is finite).  If $k=2\Nb-\Nc$ is an integer satisfying
$0\leq k \leq \Nc-1$, the integral diverges logarithmically.

(iii) If some of the $\Nc-k$ large integration variables are not
``independent'', the possible degree of divergence is reduced compared
to \eqref{eq:k-contr-U} since the effective number of large
integration variables is reduced and some differences $x_i-x_j$ stay
finite.

We therefore conclude
\begin{align}\label{eq:min-contr-U}
  \int_{V(\gamma,\mathfrak{u}(\Nc))}dH\,\det(1+H^2)^{-\Nb}\propto
  \gamma^{\min\left\{\frac12(k-\Nb)^2-\frac12(\Nc-\Nb)^2\right\}_{0\leq
      k \leq \Nc}}\,,
\end{align}
unless the minimum in the exponent on the RHS equals zero and the
corresponding $k$ satisfies $k\neq \Nc$, in which case the integral
diverges like $\log(\gamma)$.  The minimal exponent in
\eqref{eq:min-contr-U} is obtained for $k=\min\{[\Nb],\Nc\}$, where
$[\Nb]$ denotes the integer closest to $\Nb$. Hence
\begin{align}\label{eq:Uleading}
  \int_{V(\gamma,\mathfrak{u}(\Nc))}dH\,\det(1+H^2)^{-\Nb}\propto
  \begin{cases}
    \gamma^{\frac12([\Nb]-\Nb)^2-\frac12(\Nc-\Nb)^2} & \text{for }
    \Nb<\Nc-\frac12\,,\\
    \log(\gamma) & \text{for } \Nb=\Nc-\frac12\,, \\
    \gamma^0     & \text{for } \Nb>\Nc-\frac12\,,
  \end{cases}
\end{align}
where the logarithmic divergence for $\Nb=\Nc-\frac12$ results from
$k=\Nc-1$, i.e., an integration domain where one eigenvalue is ``large''
and all others remain ``finite''.

The result \eqref{eq:Uleading}, which was obtained from simple power
counting, could potentially be invalid.  After we expand the
Vandermonde determinant in \eqref{eq:Uintegral}, we obtain a sum of
factorized integrals, all of the $\U(1)$
type~\eqref{eq:cbarU1}. Cancellations in this sum could make the
coefficient of the leading-order contribution to the integral vanish.
We have explicitly checked for a range of values for $\Nc$ that this
does not happen.  This confirms the validity of the simple
power-counting arguments.

In complete analogy to the $\U(1)$ example, we expand the integrand in
the integral representation~\eqref{eq:cbar} of $\bar
c_\lambda(\gamma)$ in powers of $\gamma$. From \eqref{eq:UH},
\eqref{eq:dU}, and \eqref{eq:chiU} we obtain
\begin{align}\label{eq:bu}
  \bar c_\lambda(\gamma)=d_\lambda \int_{V(\gamma,\mathfrak{u}(\Nc)} dH \,\det(1+H^2)^{-\Nb}
  \biggl(1 +\sum_{m=1}^\infty \gamma^m b_{\lambda,m}(H)\biggr),
\end{align}
where the functions $b_{\lambda,m}(H)$ are of order $H^{2m}$ and
depend only on the eigenvalues $x_i$ of $H$. They do not depend on
$\gamma$. A term contributing to $b_{\lambda,m}(H)$ is, e.g., given by
$\bigl( \tr \frac{H^4}{1+H^2}\bigr)^m$, resulting from the expansion
\eqref{eq:UH}.  To determine the asymptotic behavior of
$\gamma^m \int_{V(\gamma,\mathfrak{u}(\Nc)} dH \,\det(1+H^2)^{-\Nb}
b_{\lambda,m}(H)$ at leading order as $\gamma \to 0$, we again analyze
the different integration domains discussed above and include an
additional factor of $(\gamma x_{Nc}^2)^m$ in \eqref{eq:Uintegral}
($x_{\Nc}$ always corresponds to a ``large'' variable in (ii) above
and reflects the fact that the $b_{\lambda,m}(H)$ are of order
$H^{2m}$).  For $m>0$ we obtain
\begin{align}\label{eq:Usub}
  \gamma^m \int_{V(\gamma,\mathfrak{u}(\Nc))}dH\,\det(1+H^2)^{-\Nb} b_{\lambda,m}(H)
  \hspace{6cm} & \cr\propto
  \begin{cases}
    \gamma^{\frac12([\Nb]-\Nb)^2-\frac12(\Nc-\Nb)^2} & \text{for }
    0<\Nb<\Nc-\frac12 \,,\\
    \gamma^{\Nb-(\Nc-\frac12)} & \text{for }
    \Nc-\frac12 \leq \Nb < \Nc+m-\frac12\,,\\
    \gamma^m\log(\gamma) & \text{for } \Nb=\Nc+m-\frac12\,, \\
    \gamma^m & \text{for } \Nc+m-\frac12<\Nb\,.
  \end{cases}
\end{align}
Note that for $m=0$ \eqref{eq:Usub} reduces to \eqref{eq:Uleading} if
we define $b_{\lambda,0}(H)=1$.  We can now analyze the dependence on
$\Nb$ of the series expansion of $\bar c_\lambda$ in powers of
$\gamma$.

For $\Nb>\Nc-\frac12$, the integral~\eqref{eq:Usub} is finite for all
$m\geq 0$. While \eqref{eq:Usub} with $m=0$ results in a contribution
of order $\gamma^0$ in the expansion of $\bar c_\lambda$, all other
$m\geq 1$ lead to contributions that vanish as $\gamma \to 0$ (terms
of order $\gamma^{\Nb-(\Nc-\frac12)}$, $\gamma^m \log \gamma$, or
$\gamma^m$).

For $\Nb=\Nc-\frac12$, the integral~\eqref{eq:Usub} diverges
logarithmically for $m=0$, while $m\geq 1$ leads to finite terms
($\gamma^{\Nb-(\Nc-\frac12)}=\gamma^0$), i.e., the contribution of
$m=0$ still dominates.

From \eqref{eq:limits} we then immediately obtain the straightforward
generalization of \eqref{eq:U1continuum},
\begin{align}\label{eq:UNcontinuum}
  \lim_{\alpha\to 1}c_\lambda(\alpha)=\lim_{\gamma\to 0} \frac{\bar
    c_\lambda(\gamma)}{\bar
    c_0(\gamma)}=\frac{d_\lambda}{d_0}=d_\lambda \qquad
  \text{for } \Nb\geq \Nc-\frac12\,.
\end{align}

For $\Nb<\Nc-\frac12$, the integral~\eqref{eq:Usub} leads to identical
leading-order divergences proportional to
$\gamma^{\frac12([\Nb]-\Nb)^2-\frac12(\Nc-\Nb)^2}$ for all $m$. In
this case, $\lim_{\alpha\to 1}c_\lambda(\alpha)$ has a non-trivial
dependence on $\lambda$ and therefore differs from $d_\lambda$
generically.

\subsubsection[$G=\SU(\Nc)$]{\boldmath $G=\SU(\Nc)$}
\label{sec:SUNdelta}

We again transform to the eigenvalue representation
$H=V\diag(x_j)V^\dagger$.  Since $H$ is now traceless we need to
incorporate the condition $\sum_jx_j=0$ in the integral.  This leads
to an additional $\delta$-function,
\begin{align}\label{eq:SUintegral}
  &\int_{V(\gamma,\mathfrak{su}(\Nc))}dH\,\det(1+H^2)^{-\Nb}\cr
  &\quad\propto
  \int_{-\pi/\sqrt\gamma}^{\pi/\sqrt\gamma}\Big(\prod_{j=1}^{\Nc}dx_j\Big)
  \delta\Big(\sum_{j=1}^{\Nc}x_j\Big) \Big(\prod_{j<k}(x_j-x_k)^2\Big)
  \Big(\prod_{j=1}^{\Nc}(1+x_j^2)^{-\Nb}\Big)\cr &\quad=
  \int_{-\pi/\sqrt\gamma}^{\pi/\sqrt\gamma}\Big(\prod_{j=1}^{\Nc-1}dx_j\Big)
  \Big(\prod_{j<k}(x_j-x_k)^2\Big)
  \Big(\prod_{j=1}^{\Nc}(1+x_j^2)^{-\Nb}\Big)
  \Bigg|_{x_{\Nc}=-\sum_{\ell=1}^{\Nc-1} x_\ell}\,.
\end{align}
In analogy with the previous subsection, we now evaluate the
integral~\eqref{eq:SUintegral} with an additional factor of
$x_{\Nc-1}^{2m}$ ($m\geq 0$) in the integrand to determine the
equivalent of \eqref{eq:Usub} for the $\SU(\Nc)$ case. To this end, we
again proceed by splitting the integral into integrals over subdomains
and analyzing their asymptotic behavior by power counting:

(i) When all $x_j$ are ``finite'' the integral is finite, i.e., of
order $\gamma^0$.

(ii) We choose an integer $k$ with $0\le k\le \Nc-2$ and consider
domains where $k$ variables ($x_i$, $1\leq i \leq k$) stay ``finite''
and $\Nc-k-1$ variables ($x_i$, $k+1\leq i \leq \Nc-1$) are ``large''.
To obtain the largest possible degree of divergence, we have chosen
the ``large'' variables such that the factor of $x_{\Nc-1}^{2m}$
always corresponds to a ``large'' variable. Furthermore, we take all
large variables to be ``independent'' (in the same sense as before,
except that we also require $x_{\Nc}=-\sum_{i=1}^{\Nc-1}x_i$ and all
differences $x_i-x_{\Nc}$ to be ``large''). Then the leading-order
contribution of such a domain to the integral is proportional to
\begin{align}\label{eq:k-contr-SU}
  \left( \frac 1{\sqrt\gamma}\right)^{\Nc-k-1+2\left[\binom\Nc2 -
  \binom k 2 \right]-2\Nb(\Nc-k)+2m}
  =\gamma^{\frac12(k-\Nb)^2-\frac12(\Nc-\Nb)^2+\frac12-m}\,,
\end{align}
provided the exponent on the RHS is negative (otherwise, the integral
is finite). If the exponent vanishes, the
integral diverges logarithmically.

(iii) If some of the $\Nc-k-1$ ``large'' integration variables are not
``independent'' such that some differences $x_i-x_j$ stay finite, the
possible degree of divergence is reduced compared to
\eqref{eq:k-contr-SU}.

(iv) If some of the $\Nc-k-1$ ``large'' integration variables are not
``independent'' such that $\sum_{j=1}^{\Nc-1}x_j$ stays finite (this
requires $k\leq \Nc-3$), the number of ``large'' integration variables
is effectively reduced by one. After a suitable change of variables we
get a contribution given by \eqref{eq:k-contr-SU} with $k$ replaced by
$k+1$.

We now proceed in analogy with \eqref{eq:bu} and replace the
coefficient functions by $\tilde b_{\lambda,m}(H)$ appropriate for
$\mathfrak{su}(\Nc)$.  Provided that leading-order contributions do
not vanish accidentally, we end up with
\begin{align}
  \gamma^m \int_{V(\gamma,\mathfrak{su}(\Nc))}dH\,\det(1+H^2)^{-\Nb} \tilde b_{\lambda,m}(H)
  \propto \gamma^{\min\left\{m,
      \left\{\frac12(k-\Nb)^2-\frac12(\Nc-\Nb)^2+\frac12\right\}_{0\leq
        k \leq \Nc-2}\right\}}\,,
\end{align}
unless the minimum in the exponent on the RHS equals $m$ and
$\frac12(k-\Nb)^2-\frac12(\Nc-\Nb)^2+\frac12=m$ for some $k\leq\Nc-2$,
in which case the integral is proportional to $\gamma^m \log(\gamma)$
at leading order.  Since $\frac12(k-\Nb)^2-\frac12(\Nc-\Nb)^2+\frac12$
is minimized by $k=\min\{[\Nb],\Nc-2\}$, we obtain
\begin{align}\label{eq:SUsub}
  \gamma^m \int_{V(\gamma,\mathfrak{su}(\Nc))}dH\,\det(1+H^2)^{-\Nb} \tilde b_{\lambda,m}(H)
  \hspace{6cm} & \cr\propto
  \begin{cases}
    \gamma^{\frac12([\Nb]-\Nb)^2-\frac12(\Nc-\Nb)^2+\frac12} &
    \text{for }
    0<\Nb\leq\Nc-\frac32 \,,\\
    \gamma^{2\left(\Nb-\left(\Nc-\frac54\right)\right)} &
    \text{for }
    \Nc-\frac52 \leq \Nb < \Nc-\frac54+\frac m2\,,\\
    \gamma^m\log(\gamma) & \text{for } \Nb=\Nc-\frac54+\frac m2\,, \\
    \gamma^m & \text{for } \Nc-\frac54+\frac m2<\Nb\,.
  \end{cases}
\end{align}
In complete analogy to the $\U(\Nc)$ case, see the discussion below
\eqref{eq:Usub} and compare \eqref{eq:SUsub} to \eqref{eq:Usub}, we
obtain
\begin{align}\label{eq:SUNcontinuum}
  \lim_{\alpha\to1 }c_\lambda(\alpha) = d_\lambda
  \qquad\text{for } \Nb\geq \Nc-\frac 54\,,
\end{align}
while the limit will generically differ from $d_\lambda$ for
$\Nb<\Nc-\frac 54$.

\subsubsection[Bounds on $\Nb$]{\boldmath Bounds on $\Nb$}

In conclusion, the necessary and sufficient condition for the weight
function~\eqref{eq:weight} to reduce to a Dirac $\delta$-function on
the group manifold in the $\alpha\to1$ limit is given by
\begin{align}
  \label{eq:delta}
  \Nb\ge
  \begin{cases}
    \Nc-\frac12 & \text{for }G=\U(\Nc)\,,\\
    \Nc-\frac54 & \text{for }G=\SU(\Nc)\,.
  \end{cases}
\end{align}
These bounds have been verified through extensive numerical
simulations.

As discussed in some detail in \cite{Budczies:2003za}, when the
statistical weight in the partition function approaches a product of
$\delta$-functions for the plaquette variables, correlation lengths
diverge and we expect the lattice theory to converge to a continuum
limit.

\subsection{Nature of the continuum limit}
\label{sec:equiv}

In the previous section we have found that the theory defined by the
weight function~\eqref{eq:weight} admits a continuum limit if the
bounds~\eqref{eq:delta} are satisfied.  To investigate the nature of
this continuum limit, we now determine the next-to-leading-order (NLO)
terms in the expansion of $c_\lambda(\alpha)$ about $\alpha=1$, i.e.,
NLO corrections to \eqref{eq:UNcontinuum} and~\eqref{eq:SUNcontinuum}.

\subsubsection[NLO terms for $G=\U(\Nc)$]{\boldmath NLO terms for
  $G=\U(\Nc)$}
\label{sec:NLO-U}

For $G=\U(\Nc)$ and $\Nb>\Nc-\frac12$, the
integral \eqref{eq:Usub} is finite as
$\gamma \to 0$. For $\Nb\geq \Nc+\frac12$, we furthermore see that the
result for $m=1$ dominates over all terms with $m>1$. This means that
the NLO term in $c_\lambda(\alpha)$ is exclusively determined by the
first-order term of the expansion of the integrand in
\eqref{eq:bu}.
From the expansion~\eqref{eq:chiU} of $\chi_\lambda$ we thus obtain
for $\Nb>\Nc+\frac12$
\begin{align}\label{eq:U-NLO}
  c_\lambda(\alpha)=\frac{\bar c_\lambda(\gamma)}{\bar c_0(\gamma)}=
  d_\lambda \left\{1 -\frac{\gamma}2\left[
      \left(\frac{q(\lambda)^2}{\Nc^2}-\frac{A(\lambda)}{\Nc}\right)\!\left\langle(\tr
        H)^2\right\rangle_{\mathfrak u(\Nc)} +A(\lambda) \left\langle
        \tr H^2 \right\rangle_{\mathfrak u(\Nc)} \right]+o(\gamma)
  \right\}
\end{align}
with
\begin{align}
  \label{eq:fg}
  \left\langle f(H) \right\rangle_{\mathfrak g} =
  \frac{\int_{\mathfrak g} dH\,\det(1+H^2)^{-\Nb}\,
    f(H)}{\int_{\mathfrak g} dH\,\det(1+H^2)^{-\Nb}}
\end{align}
and $o(\gamma)$ refers to the little-o notation.  Note that subleading
corrections to $\int_V dH \det(1+H^2)^{-\Nb} \propto \gamma^0$ result
only in terms of order $o(\gamma)$ in the ratio
$\bar c_\lambda / \bar c_0$ since the integral does not depend on
$\lambda$.  Furthermore, contributions from $\lambda$-independent
terms in $b_{\lambda,1}(H)$ cancel at order $\gamma$ in
\eqref{eq:U-NLO}.

For $\Nb=\Nc+\frac12$, $\gamma$ on the RHS of \eqref{eq:U-NLO} has to
be replaced by $\gamma \log(\gamma)$, and the $\gamma\to0$ limit
implicit in \eqref{eq:fg}, i.e.,
$\lim_{\gamma\to0}V(\gamma,{\mathfrak g})=\mathfrak g$, has to be
taken more carefully, see \eqref{eq:Uleading} and \eqref{eq:Usub}.

For $\Nc-\frac12 \leq \Nb < \Nc+\frac12$, \eqref{eq:Usub} yields the
same leading-order term for all $m\geq 1$, proportional to
$\gamma^{\Nb-(\Nc-\frac12)}$, which implies that the NLO term in
$c_\lambda(\alpha)$ is not simply determined by the term of order
$\gamma$ in the expansion \eqref{eq:chiU} of the character
$\chi_\lambda$.

For $\U(1)$, the coefficient $c_\lambda$ can be calculated
analytically for $\Nb \in \mathbbm N$, confirming the results derived
above (see appendix~\ref{App:u1}).  Our result~\eqref{eq:U-NLO} is in
agreement\footnote{ From \eqref{eq:mtwid} and $\alpha\BZ=m\BZ^{-4}$ we
  obtain $\gamma=(1-\alpha\BZ)^2/\alpha\BZ=(1-\alpha\BZ)^2+\ldots$
  Note that $X$ in \cite[eq.~(25)]{Budczies:2003za} corresponds to our
  $iH$.  } with \cite[eq.~(25)]{Budczies:2003za}, which was derived
for integer $\Nb\geq\Nc+1$.

\subsubsection[NLO terms for $G=\SU(\Nc)$]{\boldmath NLO terms for
  $G=\SU(\Nc)$}
\label{sec:SUN-NLO}

In complete analogy to the $\U(\Nc)$ case, see \eqref{eq:Usub}
compared to \eqref{eq:SUsub}, we obtain from \eqref{eq:chiSU} for
$\Nb>\Nc-\frac34$
\begin{align}\label{eq:SU-NLO}
  c_\lambda(\alpha)=\frac{\bar c_\lambda(\gamma)}{\bar c_0(\gamma)}=
  d_\lambda \left[1 - \gamma
    \,\frac{C_2^{\SU}(\lambda)}{\Nc^2-1}\left\langle \tr
      H^2\right\rangle_{\mathfrak {su}(\Nc)} + o(\gamma) \right].
\end{align}
For $\Nb=\Nc-\frac34$, the comment made after \eqref{eq:U-NLO}
applies.  For $\Nc-\frac54 \leq \Nb < \Nc -\frac34$, all terms in the
expansion of the integrand contribute to the NLO term in
\eqref{eq:SU-NLO}, resulting in a more complicated dependence on
$\lambda$ compared to \eqref{eq:SU-NLO}.  For $\SU(2)$, the
coefficient $c_\lambda$ can be calculated analytically for
$2\Nb \in \mathbbm N$, confirming these results (see
appendix~\ref{App:su2}).

Assuming large $\Nb$, we now perform a saddle-point expansion about
the trivial saddle at $H=0$ and obtain
\begin{align}\label{eq:trH^2}
  \Nb \left\langle \tr H^2 \right\rangle_{\mathfrak {su}(\Nc)} &=
  \frac12 (\Nc^2-1)\biggl[ 1
  +\left(\Nc-\frac{3}{2\Nc}\right)\frac1{\Nb}
  +\left(\Nc^2-\frac{15}4+\frac{21}{4\Nc^2}\right)\frac1{\Nb^2} \cr &
  \quad +\left(\Nc^3-\frac{27
      \Nc}4+\frac{93}{4\Nc}-\frac{267}{8\Nc^3}\right)\frac1{\Nb^3}
  +\mathcal O\left(\Nb^{-4}\right) \biggr]\,,
\end{align}
where the effective expansion parameter appears to be $\Nc/\Nb$. For
$\SU(2)$, the exact result is given by
\begin{align}\label{Eq:TrHsqSU2}
  \left.  \frac{\frac12 (\Nc^2-1)}{\Nb \left\langle \Tr H^2
      \right\rangle_{\mathfrak {su}(\Nc)}}
  \right\rvert_{\Nc=2}=1-\frac{5}{4 \Nb}\,.
\end{align}
For small $\Nc$, it is more convenient to use the eigenvalue
parameterization of $H\in \mathfrak{su}(\Nc)$ (instead of the
parameterization as a linear combination of generators) for the
saddle-point approximation since the computation of higher-order terms
can then be automated easily. In this way, we obtain
\begin{align}
  \left.  \frac{\frac12 (\Nc^2-1)}{\Nb \left\langle \Tr H^2
      \right\rangle_{\mathfrak {su}(\Nc)}} \right\rvert_{\Nc=3}&= 1
  -\frac{5}{2 \Nb} +\frac{5}{12 \Nb^2} +\frac{5}{18 \Nb^3}
  -\frac{95}{432 \Nb^4} -\frac{485}{2592 \Nb^5} +\frac{12715}{7776
    \Nb^6} \cr&\quad +\frac{127445}{93312 \Nb^7}
  -\frac{4267895}{559872 \Nb^8} +\frac{6392335}{839808 \Nb^9}
  +\frac{1424010605}{20155392 \Nb^{10}}
  % -\frac{25299882505}{120932352 \Nb^{11}}
  % -\frac{147326509285}{362797056 \Nb^{12}}
  % +\frac{19557429791665}{4353564672 \Nb^{13}}
  % -\frac{152261937210595}{26121388032 \Nb^{14}}
  % -\frac{2959266203415895}{39182082048 \Nb^{15}}
  % +\frac{384505058433018505}{940369969152 \Nb^{16}}
  % +\frac{2716331623962402115}{5642219814912 \Nb^{17}}
  % -\frac{243191484574418734085}{16926659444736 \Nb^{18}}
  % +\frac{9426134116462724754845}{203119913336832 \Nb^{19}}
  % +\frac{391711425530246583832705}{1218719480020992 \Nb^{20}}
  +\ldots
  \\
  \left.  \frac{\frac12 (\Nc^2-1)}{\Nb \left\langle \Tr H^2
      \right\rangle_{\mathfrak {su}(\Nc)}} \right\rvert_{\Nc=4}&= 1
  -\frac{29}{8 \Nb} +\frac{9}{16 \Nb^2} +\frac{81}{64 \Nb^3}
  +\frac{207}{128 \Nb^4} -\frac{27}{64 \Nb^5} -\frac{14787}{4096
    \Nb^6}
  % +\frac{262305}{16384 \Nb^7} +\frac{6390477}{65536 \Nb^8}
  % +\frac{11615643}{262144 \Nb^9} -\frac{840143331}{1048576 \Nb^{10}}
  % +\frac{1685407635}{4194304 \Nb^{11}} +\frac{74632592763}{4194304
  % \Nb^{12}}
  +\ldots
\end{align}

\subsubsection[Character expansion for $\SU(\Nc)$ Wilson plaquette
action]{\boldmath Character expansion for $\SU(\Nc)$ Wilson plaquette
  action}
\label{Sec:charWilson}

To determine the nature of the continuum limit of the boson-induced
lattice gauge theory, we compare \eqref{eq:SU-NLO} to the
corresponding expansion of the familiar $\SU(\Nc)$ Wilson plaquette
action, see \eqref{Eq:SWilson} below.  In analogy to
\eqref{eq:charexp} we expand for $U\in \SU(\Nc)$
\begin{align}
  \frac1{Z_W}e^{\frac1{\gw^2} \Tr(U+U^\dagger-2)}=\sum_\lambda
  c^{(W)}_\lambda(\gw^2) \chi_\lambda(U)
\end{align}
with normalization factor $Z_W$ defined in the obvious manner.  Using
the parameterization $U=e^{i\gw A}$ with $A\in\mathfrak{su}(\Nc)$ we
obtain
\begin{align}\label{Eq:cbar-linear-Wilson}
  c^{(W)}_\lambda(\gw^2) = d_\lambda \biggl( 1 - \gw^2 \frac{
      C_2^{\SU(\Nc)}(\lambda) }{\Nc^2-1} \left\langle \Tr A^2
    \right\rangle_W +\ldots \biggr)\,,
\end{align}
where
\begin{align}
  \left\langle \Tr A^2 \right\rangle_W =
  \frac{\int_{\mathfrak{su}(\Nc)} dA\, e^{-\Tr A^2} \Tr
    A^2}{\int_{\mathfrak{su}(\Nc)} dA\, e^{-\Tr A^2}}= \frac12
  \left(\Nc^2-1\right)
\end{align}
is just the leading term of $\Nb \left\langle\Tr H^2 \right\rangle$
computed in section~\ref{sec:SUN-NLO} by saddle-point approximation.
Truncating the expansions of $c_\lambda(\alpha)$ and
$c^{(W)}_\lambda(\gw^2)$ after the NLO term, both weight factors
become equivalent to the heat-kernel weight
\begin{align}
  \omega_{\text{HK}}(U,t)=\sum_\lambda d_\lambda e^{-t\,
    C_2^{\SU(\Nc)}(\lambda)} \chi_\lambda(U)
\end{align}
with diffusion times $t\propto \gamma \propto (1-\alpha)$ and
$t\propto \gw^2$, respectively. A similar equivalence holds for
$G=\U(\Nc)$ \cite{Budczies:2003za}.

\subsubsection{Continuum limit in 2d}

In two dimensions, the heat-kernel lattice action is exactly
self-reproducing and therefore invariant under subdivision of the
lattice (Migdal's recursion \cite{Migdal:1975zg}).  Consider, e.g.,
two neighboring plaquettes $p_1$ and $p_2$, where $p_2$ is to the
right of $p_1$. If the common link variable is called $U$, the
plaquette variables are parametrized as $U_{p_1}=V_1 U W_1$ and
$U_{p_2}=U^\dagger W_2$. Then, due to character
orthogonality~\eqref{eq:char-orth},
\begin{align}
  \int_G dU\, \omega_{\text{HK}}(U_{p_1},t)
  \omega_{\text{HK}}(U_{p_2},t) =
  \omega_{\text{HK}}(U_{p_1+p_2},2t)\,,
\end{align}
where $U_{p_1+p_2}=V_1 W_2 W_1$ is the Wilson loop variable for the
boundary of the joint lattice cell $p_1+p_2$.

If we set $t = t_p a^2$, with $t_p$ (of dimension 1/area) fixed in
the continuum limit $a\to 0$, and consider a region $R$ of physical
area $A_R$ in flat spacetime, obtained by gluing together $n=A_R/a^2$
elementary plaquettes, the effective action for $U_R$ (the product of
link variables along the boundary of $R$), determined by integrating
over all internal link variables, is given by
\begin{align}\label{eq:effHK}
  \omega_{\text{HK}}(U_R,n t)= \omega_{\text{HK}}(U_R, A_R t_p)\,.
\end{align}
Since the effective action has the same functional form as the
original plaquette action and only the diffusion parameter changes
proportional to the enclosed area, taking the continuum limit is
trivial in two dimensions for the heat-kernel action.  In flat
spacetime, Wilson loop variables (corresponding to closed
non-selfintersecting curves enclosing an area $A_R$) are simply
distributed according to the distribution~\eqref{eq:effHK}.

From the effective action~\eqref{eq:effHK} for fundamental polygons,
the YM partition function on an orientable surface of genus $g$ and
dimensionless area $A$ (in suitable units) was found in
\cite{Witten:1991we} to be given by
\begin{align}
  Z_g(A)=\sum_\lambda d_\lambda^{2- 2g} e^{-A
    C_2^{\SU(\Nc)}(\lambda)}\,.
\end{align}
From \eqref{eq:SU-NLO} and~\eqref{Eq:cbar-linear-Wilson} we see that
using either the induced or the Wilson weight function instead of the
heat-kernel distribution for the elementary plaquette variables still
leads to the effective action~\eqref{eq:effHK} in the continuum limit
$a\to 0$ if we scale $\gamma\propto a^2$ and $\gw^2\propto a^2$,
respectively. This implies that the continuum limit of the induced
$\SU(\Nc)$ theory with $\Nb\geq \Nc-\frac34$ is equivalent to YM
theory in two dimensions. Similarly, we conclude from the results of
section~\ref{sec:NLO-U} that the continuum limit for $\U(\Nc)$ is
equivalent to YM for $\Nb\geq \Nc+\frac12$.

To ensure that the induced action and the Wilson action lead to the same
physics in the continuum limit in two dimensions we require
$c_\lambda(\alpha)=c_\lambda^{(W)}(g_W^2)$ to NLO and thus obtain from
\eqref{eq:SU-NLO} and \eqref{Eq:cbar-linear-Wilson}
\begin{align}\label{eq:alpha-gw-2d}
  \frac1{\gw^2}&=\frac{1}{\gamma} \frac{\left\langle \Tr A^2
    \right\rangle_W}{\left\langle \Tr H^2
    \right\rangle_{\mathfrak{su}(\Nc)}} \cr &=
  \frac{\Nb}{\gamma} \biggl( 1 -\frac{1}{2 \Nc}(2 \Nc^2-3)
  \frac1{\Nb} +\frac{3}{4 \Nc^2}(\Nc^2-4)\frac1{\Nb^2}\cr
  &\qquad\qquad+\frac{3}{4\Nc^3}(\Nc^2-4)(\Nc^2-7) \frac1{\Nb^3}
  +\ldots \biggr)\,.
\end{align}
This relation between the couplings is consistent with the more
general result obtained from perturbation theory in an arbitrary
number of dimensions, see section~\ref{sec:pert} below. In higher
dimensions, taking the continuum limit is of course more involved.

\subsubsection[Bounds on $\Nb$]{\boldmath Bounds on $\Nb$ in 2d}

In conclusion, the necessary and sufficient condition for the
continuum limit of the induced theory to be equivalent to YM theory in
two dimensions is
\begin{align}
  \label{eq:cont}
  \Nb\ge
  \begin{cases}
    \Nc+\frac12 & \text{for }G=\U(\Nc)\,,\\
    \Nc-\frac34 & \text{for }G=\SU(\Nc)\,.
  \end{cases}
\end{align}

\subsubsection{Continuum limit in 3d and 4d}
\label{Sec:cont3d4d}

Following \cite{Budczies:2003za}, we conjecture that the equivalence
with YM theory persists also in higher dimensions.  Furthermore, the
``exotic'' case $\Nb=\Nc$ for $G=\U(\Nc)$ in two dimensions, where the
continuum limit of the induced theory differs from YM theory, was
studied in great detail in \cite{Budczies:2003za}.  BZ argue that this
unusual theory of Cauchy-type is unlikely to persist in three or four
dimensions. Similarly, we expect the continuum limit of the induced
theory to be equivalent to YM theory in $d>2$ for both $G=\U(\Nc)$ and
$G=\SU(\Nc)$ whenever the continuum limit actually exists, i.e., if
\eqref{eq:delta} is satisfied. Our numerical tests support these
expectations \cite{Brandt:2014rca,BLW:II}.

\section{Perturbative matching of the couplings}
\label{sec:pert}

In the following, we consider only $G=\SU(\Nc)$ since this case
includes the gauge group of QCD and is therefore of phenomenological
interest.

\subsection{General strategy}
\label{Sec:Strategy}

Since the continuum limit is essentially trivial in two dimensions,
the relation between $1-\alpha$ and the Wilson coupling $\gw^2$ can be
obtained simply by matching the character expansions of the plaquette
weight functions, resulting in \eqref{eq:alpha-gw-2d} above. On the
other hand, in three and four dimensions the continuum limit is more
involved so that we need an alternative method to determine the
relation between the bare couplings. A natural candidate is
perturbation theory, which we will use in the following.

Ideally, one would like to expand around the continuum limit at
$\alpha=1$ for fixed $\Nb$.  However, we then encounter two problems.
First, the expansion of the logarithm in
\begin{align}
  S_I&= -\log \prod_p {\det}^{-\Nb}\left(1-\frac \alpha
    2\bigl(U_p+U_p^\dagger\bigr)\right) =\Nb \sum_p \Tr
  \log\Bigl(1-\frac \alpha 2\bigl(U_p+U_p^\dagger\bigr)\Bigr)\cr &=
  \Nb \sum_p \Tr \log\Bigl(1-\frac \alpha
    {2(1-\alpha)}\bigl(U_p+U_p^\dagger-2\bigr)\Bigr)\,,
\end{align}
where we omitted an irrelevant constant in the last step, converges
only if $\big | \frac{\alpha}{1-\alpha}(\cos\varphi-1) \big | \leq 1$
for all possible eigenvalues $e^{i \varphi}$ of $U_p$, i.e., if
$\alpha \leq \frac13$.  Second, after expanding the logarithm anyway,
we see that a saddle-point analysis of the partition function is not
possible since higher orders of $U+U^\dagger -2$ are not suppressed in
\begin{align}\label{Eq:Sind}
  S_I=-\Nb \sum_p \sum_{n=1}^\infty \frac 1 n \frac1{\gamma^n}
  \Tr \bigl(U_p+U_p^\dagger-2\bigr)^{{n}}
\end{align}
with $\gamma=2(1-\alpha)/\alpha$ as defined in \eqref{eq:gamma}, and
we would end up with non-Gaussian integrals.\footnote{ For simplicity,
  consider only a single plaquette and parametrize
  $U=e^{i\sqrt{\gamma} H}$ such that the $H^2$-term in $S_I$ has a
  coefficient of order $\gamma^0$. Then, an expansion in powers of
  $\gamma$ results in
  \begin{align}
    S_I=-\Nb\sum_{n=1}^\infty \frac{(-1)^n}n \Tr H^{2n} + \mathcal
    O(\gamma)=\Nb\log\det(1+H^2)+\mathcal O(\gamma)\,,
  \end{align}
  as expected from \eqref{eq:UH}. This means that all powers of
  $H^2$ contribute to the action at leading order in $\gamma$.  }
As a workaround, we will therefore first keep $\alpha \leq \frac13$
fixed (i.e., $\gamma\geq 4$) and take the limit
$\Nb\to\infty$, which allows for a systematic saddle-point analysis,
and then analytically continue $\gw(\alpha,\Nb)$ to small $1-\alpha$
at the end.
 
It is natural to define the coupling $\tgi$ for the induced theory in
the limit $\Nb\to\infty$ at fixed $\alpha$ as\footnote{One could of
  course choose to include subleading terms in $1/\Nb$ in the
  definition of $\tgi$, but the definition in \eqref{Eq:gI} seems to
  be the most natural choice.}
\begin{align}\label{Eq:gI}
  \frac{1}{\tgi^2}= \Nb
  \frac{\alpha}{2(1-\alpha)}=\frac{\Nb}{\gamma}
\end{align}
since the induced action in terms of this coupling $\tgi$ and the
fixed parameter $\gamma\geq 4$ reads
\begin{align}\label{Eq:S1gIm}
  S_I=-\frac{1}{\tgi^2} \biggl\{ \sum_p \Tr\bigl(U_p+U_p^\dagger
      -2\bigr)+\sum_{n=2}^\infty \sum_p \frac 1{n \gamma^{n-1}}
    \Tr\bigl(U_p+U_p^\dagger-2\bigr)^n\biggr\}\,,
\end{align}
where the first term is identical to the Wilson gauge action
\begin{align}\label{Eq:SWilson}
  S_{W}= -\frac 1{\gw^2} \sum_p \Tr\bigl(U_p+U_p^\dagger -2\bigr)
\end{align}
with coupling $\gw$ replaced by $\tgi$. All sums over $p$ are sums
over unoriented plaquettes, i.e., $p=(x,\mu<\nu)$.  Expanding
\eqref{Eq:S1gIm} around $U=\1$ in the usual manner (at fixed
$\gamma$), we observe that the induced action $S_I$ reduces to the
familiar YM action in the continuum limit $U\to\1$.

It is obvious from \eqref{Eq:S1gIm} that the induced action reproduces
the Wilson action at the lattice level (i.e., at non-zero $g$) if we
take the limit $\gamma \to \infty$ at fixed $\tgi$, corresponding to
the limit discussed in connection with \eqref{eq:BZ}. However, this is
not what we are interested in here. In the following, we keep $\gamma$
fixed and expand in powers of $\tgi$.

At fixed $\gamma$, the coupling $\tgi$ plays a role which is in
complete analogy to that of $\gw$ for the Wilson action. Parametrizing
the link variables as $U=e^{i\tgi A}$, functional integrals can be
systematically expanded in powers of $\tgi$ in a saddle-point
analysis.  A character expansion of the plaquette weight function
results in coefficients $c_\lambda$ which are identical, up to
$\mathcal O(g^2)$, to those that we obtained for the Wilson action in
section~\ref{Sec:charWilson}. This implies that the weight function
reduces to a $\delta$-function on the $\SU(\Nc)$ manifold in the limit
$\tgi\to 0$ (at fixed $\gamma$) and that the continuum limit is
equivalent to YM theory in two dimensions.
Therefore, keeping $\gamma$ fixed and taking $\tgi\to 0$, we expect
the induced theory to describe YM theory in the continuum for all
$\gamma \geq 4$ also in three and four dimensions (with a dependence
on $\gamma$ occurring, e.g., in the ratio
$\Lambda_{\text{lat}}/\Lambda_{\text{cont}}$ of the $\Lambda$ parameters).

Moreover, expanding the action and the partition function in $\tgi$,
the same power-counting arguments apply as in the familiar
Wilson case. In particular, using the back\-ground-field technique, we
only have to expand the action to quadratic order in the quantum
fields to compute the effective two-point function for the background
fields at one-loop order (which determines the ratio of the $\Lambda$
parameters).

Our aim in this part of the paper is to calculate the relation between
the couplings
\begin{align}\label{Eq:gWgI}
  \frac 1{\gw^2}=\frac{1}{\tgi^2}\left(1+c_1(\gamma) \tgi^2 +
    c_2(\gamma)\tgi^4+\ldots\right).
\end{align}
We will see in sections~\ref{Sec:Gamma1loop} and \ref{Sec:Gamma2loop}
below that
\begin{align}
  c_1(\gamma) & = \frac{c_{1,-1}}{\gamma}\,, \label{Eq:c1} \\
  c_2(\gamma) & = \frac{c_{2,-2}}{\gamma^2} +
  \frac{c_{2,-1}}{\gamma}\,.  \label{Eq:c2}
\end{align}
The one-loop coefficient $c_1(\gamma)$ in \eqref{Eq:gWgI}
directly determines the ratio of the $\Lambda$ parameters,
\begin{align}
  \frac{\Lambda_I(\gamma)}{\Lambda_W} = \exp\left(\frac{c_1(\gamma)}{
      2 \beta_0}\right),\qquad \beta_0 = \frac{11 \Nc}{48 \pi^2}\,.
\end{align}
It will turn out, see \eqref{Eq:c1result} below, that $c_1(\gamma)$ is
always negative.  The two-loop coefficient $c_2(\gamma)$ determines
the first non-universal coefficient $\beta_2$ in the $\beta$ function.

When we write $1/\gw^2$ as a function of $\gamma$ and $\Nb$, it turns
out that only simple poles in $\gamma$ appear (also when we extend
\eqref{Eq:gWgI} to higher orders in $\tgi$, see
section~\ref{Sec:PowerCounting} below), i.e.,
$c_n(\gamma)\propto \gamma^{-n}$ for $\gamma\to 0$.  Replacing
$\tgi^2$ by $\gamma/\Nb$ in \eqref{Eq:gWgI}, we may formally regard
$1/\gw^2$ as a series in $\gamma$ with coefficients depending on $\Nb$
(although the relation is strictly valid only in the limit
$\Nb\to\infty$ at fixed $\gamma \geq 4$).  Assuming that we can
analytically continue \eqref{Eq:gWgI} to small
$\gamma=2(1-\alpha)/\alpha$, we obtain
\begin{align}\label{eq:match}
  \frac{1}{\gw^2}=\frac{d_0(\Nb)}{\gamma}+d_1(\Nb)+\mathcal O(\gamma)
\end{align}
with
\begin{align}
  d_0(\Nb) &= \Nb+c_{1,-1}+\frac{c_{2,-2}}{\Nb}+\mathcal
  O\left(\Nb^{-2} \right),
  \label{Eq:d0pert}\\
  d_1(\Nb) &= \frac{c_{2,-1}}{\Nb}+\mathcal O\left(\Nb^{-2}
  \right). \label{Eq:d1pert}
\end{align}
For the limit $\gamma\to0$ (i.e., $\alpha \to 1$) at fixed $\Nb$, a
natural definition of the coupling is thus given by\footnote{Again, it
  is possible to include subleading terms in $\gamma$ in the
  definition of $\gi$, effectively changing the coefficients $d_j$
  ($j\geq1$).}
\begin{align}\label{Eq:gItilde}
  \frac1{\gi^2}=\frac{d_0(\Nb)}{\gamma}
\end{align}
so that
\begin{align}\label{Eq:gWgItilde}
  \frac1{\gw^2}=\frac{1}{\gi^2}\left(1+d_1(\Nb) \gi^2+\ldots\right).
\end{align}
In the following, we will calculate $c_{1,-1}$ and $c_{2,-2}$ using
the background-field technique. The computation of the remaining
two-loop coefficient $c_{2,-1}$ is considerably more involved and
therefore left for future work.

\subsection{Background-field calculation}

\subsubsection{Effective action}

The background-field technique was introduced in \cite{DeWitt:1967ub}.
Following \cite{weinbergII,Dashen:1980vm,Hasenfratz:1981tw}, we define
the effective action
\begin{align}\label{Eq:Gamma}
  e^{-\Gamma[A]} \propto \int_{\text{1PI}}[Dq]\, e^{-S[A,q]}\,,
\end{align}
where $A$ is the background field, $q$ is the quantum field, and the
path integral is over one-particle irreducible graphs with
an arbitrary number of external lines. Here, $A$ is not required to
satisfy the YM field equations. We will expand only to quadratic order
in $A$ since this is sufficient to determine the relation of the
couplings.  In the expansion of the action $S$, terms linear in $q$
can be omitted since they do not contribute to 1PI diagrams.

The gauge-fixing procedure for the induced theory can be taken over
one-to-one from the Wilson case.  It is convenient to use the
background-field gauge since the effective action $\Gamma[A]$ is then
invariant under formal background-field transformations (resulting in
constraints on renormalization parameters).  We argue below that
diagrams with ghost loops cannot contribute to $c_{1,-1}$ and
$c_{2,-2}$. Integrals over ghost fields are therefore already omitted
in \eqref{Eq:Gamma}.  Also, we can ignore the renormalization of the
gauge-fixing parameter since (i) we will not compute $c_{2,-1}$ and
(ii) the coefficient $c_{2,-2}$ is determined exclusively by the
two-point function of the background field, see
section~\ref{Sec:PowerCounting}.

The relation between the couplings $\gw$ and $\tgi$ is obtained by
requiring $\Gamma_I[A] = \Gamma_W[A]$ in the continuum limit $g\to 0$.
To compute the effective action, we expand $S[A,q]$ (including
gauge-fixing terms) in powers of the quantum field $q$, separate the
classical piece (i.e., terms independent of $q$) and the free part
(i.e., terms of order $A^0q^2$) of the action from interaction terms
(i.e., all other terms) and compute their (one-particle irreducible
connected) expectation values w.r.t.~the free action, 
\begin{align}\label{Eq:GammaExpanded}
  \Gamma[A] &= S_{\text{cl}}[A] - \sum_{k=1}^\infty \frac 1{k!}
  \Bigl\langle \left(- S_{\text{int}}[A,q]\right)^k
  \Bigr\rangle_{\text{1PI-C}}\,.
\end{align}
Since we are only interested in the two-point function for the
background field it is sufficient to calculate expectation values of
$S_{\text{int}}^k$ at order $A^2$.
Requiring $\Gamma_I[A] = \Gamma_W[A]$ in the continuum limit will
result in an equation of the form
\begin{align}\label{Eq:gWgI-full}
  \frac 1{\gw^2} + c_1^{(W)} + c_2^{(W)} \gw^2 + \mathcal O(\gw^4) =
  \frac{1}{\tgi^2}+ \bigl(c_1^{(W)}+c_1(\gamma)\bigr) +
  \bigl(c_2^{(W)}+c_2(\gamma)\bigr) \tgi^2 + \mathcal O(\tgi^4)\,,
\end{align}
where we have split the coefficients for the induced theory into
contributions that originate exclusively from the Wilson part of the
action and terms that depend on $\gamma$. Since
$\gw^2=\tgi^2+\mathcal O(\tgi^4)$ we end up with
\eqref{Eq:gWgI}. Obviously, $c_{1}^{(W)}$ drops out of
\eqref{Eq:gWgI-full}, and $c_{2}^{(W)}$ drops out in $\mathcal O(g^2)$
and therefore does not need to be computed explicitly.

\subsubsection{Expansion of the gauge action}

We parametrize the link variables as
\cite{Dashen:1980vm,Hasenfratz:1981tw}\footnote{Note that
  \eqref{Eq:UqA} corresponds to \cite[eq.~(4)]{Hasenfratz:1981tw} with
  $q_\mu=-\alpha_\mu$ and $A_\mu=-W_\mu$ since
  $U_\mu(x)=U(x,x+\mu)^\dagger$.}
\begin{align}\label{Eq:UqA}
  U_\mu(x)=U^{(0)}_\mu(x) e^{i a g q_\mu(x)} \,,\qquad U^{(0)}_\mu(x)
  = e^{i a A_\mu (x)}\,,
\end{align}
where we imply $g=\tgi$ or $g=\gw$ for the induced action and the Wilson
action, respectively.  Since we need to expand the gauge action only
to quadratic order in $A$, we write
\begin{align}\label{eq:SI}
  S_I = \left. S_W \right|_{\gw=\tgi} + \sum_{n=2}^\infty
  \left(S_I^{(n,0)}+S_I^{(n,1)}+S_I^{(n,2)}+\mathcal O(A^3) \right),
\end{align}
where $S_I^{(n,k)}$ includes all $\mathcal O(A^k)$ terms resulting
from $\Tr(U_p+U_p^\dagger -2)^n$ in the sum over $n$ in
\eqref{Eq:S1gIm}. Defining
\begin{align}
  q_{\mu\nu}(x)&= q_\mu(x)+q_\nu(x+\mu)-q_\mu(x+\nu)-q_\nu(x)\,, \label{Eq:qmn} \\
  A_{\mu\nu}(x)&=
  A_\mu(x)+A_\nu(x+\mu)-A_\mu(x+\nu)-A_\nu(x) \label{Eq:Amn}
\end{align}
we obtain to leading order in the quantum field (see
appendix~\ref{App:ExplicitExpansion} for details)
\begin{align}
  S_I^{(n,0)} & = (-1)^{n+1}
  \frac{a^{2n}g^{2n-2}}{\gamma^{n-1}} \sum_{x,\mu,\nu}
  \frac{1}{2n} \Tr\left[ q_{\mu\nu}(x)^{2n}+
    \mathcal O \left(gq^{2n+1}\right)\right] , \label{Eq:SInk0}\\
  S_I^{(n,1)} & = (-1)^{n+1}
  \frac{a^{2n}g^{2n-3}}{\gamma^{n-1}} \sum_{x,\mu,\nu}
  \Tr\left[ A_{\mu\nu}(x) q_{\mu\nu}(x)^{2n-1}+
    \mathcal O \left(g A q^{2n} \right)\right], \label{Eq:SInk1}\\
  S_I^{(n,2)} & = (-1)^{n+1}
  \frac{a^{2n}g^{2n-4}}{\gamma^{n-1}} \sum_{x,\mu,\nu}
  \Tr\biggl[ \frac 12 \left(A_{\mu\nu}(x)q_{\mu\nu}(x)^{n-1}\right)^2
  \cr &\qquad\qquad + \sum_{m=0}^{n-2} A_{\mu\nu}(x) q_{\mu\nu}(x)^m
  A_{\mu\nu}(x) q_{\mu\nu}(x)^{2n -m -2} + \mathcal O \left(g A^2
    q^{2n-1}\right) \biggr]\,. \label{Eq:SInk2}
\end{align}

\subsubsection{Gauge fixing and free action for the quantum field}
\label{Sec:FreeAction}

For the expansion of $S_W[U]$ in terms of $A$ and $q$ using the
parameterization~\eqref{Eq:UqA}, as well as for the gauge-fixing
procedure, we can use the results of
\cite{Dashen:1980vm,Hasenfratz:1981tw}.  Since we do not have to
compute $c_{1,2}^{(W)}$ in \eqref{Eq:gWgI-full} to determine the
relation between $\gw$ and $\tgi$, all we need here is the free
(gauge-fixed) action for the quantum field $q$ to quadratic order.

The gauge-fixing term in background-field Feynman gauge is given by
\begin{align}
  S_{\text{gf}}=a^4 \sum_x \Tr\biggl( \sum_{\mu} \bar D^{(0)}_\mu q_\mu
  \biggr)^2
\end{align}
with the lattice covariant derivative (involving only the background
field)
\begin{align}
  \bar D^{(0)}_\mu q_\nu(x) = \frac1a\left( U^{(0)}_\mu(x-\mu)
    q_\nu(x-\mu) U^{(0)\dagger}_\mu(x-\mu) - q_\nu(x) \right).
\end{align}
The free action for the quantum field $q$ is obtained by combining
$S_{\text{gf}}|_{A=0}$ with the terms of order $A^0q^2$ in the gauge
action $S_I$.  The latter are obtained from $S_I^{(1,0)}$, which is
defined in the sentence following \eqref{eq:SI} and given explicitly
in \eqref{Eq:SInk0}.  $S_I^{(1,0)}$ is part of $S_W$ on the RHS of
\eqref{eq:SI}.  This means that the free action is identical for the
induced and the Wilson gauge action. It is found to be given by
\begin{align}
  S_f = a^4 \sum_{x,\mu,\nu} \Tr \left( \Delta_\mu q_\nu(x)\right)^2 =
  a^4 \sum_{x,\nu} \Tr \left( q_\nu(x) \square q_\nu(x) \right)
\end{align}
with lattice derivatives
\begin{align}
  \Delta_\mu f(x) & = a^{-1}\left( f(x+\mu) - f(x) \right),\\
  \bar \Delta_\mu f(x) & = a^{-1}\left( f(x-\mu) - f(x) \right), \\
  \square & = \sum_\mu \bar \Delta_\mu \Delta_\mu\,.
\end{align}
Writing $q_\mu$ as a linear combination of $\SU(\Nc)$ generators,
$q_\mu=\sum_{b=1}^{\Nc^2-1} q_\mu^b t_b$, we obtain with the normalization
condition~\eqref{Eq:GenNorm}
\begin{align}
  S_f = \frac {a^4}2 \sum_{x,\mu,b} q_\mu^b(x) \square q_\mu^b(x)\,,
\end{align}
which is just the free action of a collection of $d(\Nc^2-1)$
independent massless scalar fields. Here, $d$ denotes the number of
Euclidean spacetime dimensions.  The free propagator is therefore
given by
\begin{align}\label{Eq:qPropagator}
  D_{\mu\nu}^{ab}(x,y)= \bigl\langle q_\mu^a(x) q_\nu^b(y)
  \bigr\rangle = \delta_{ab}\delta_{\mu\nu} D(x-y)
\end{align}
with the standard lattice propagator for a massless scalar field
\begin{align}\label{Eq:Propagator}
  D(x-y) = \int_{-\pi/a}^{\pi/a} \frac{d^dp}{(2\pi)^d} e^{ip(x-y)}
  \frac{a^{d-2}}{\sum_\mu 2\left(1- \cos\left( a
        p_\mu\right)\right)}\,.
\end{align}
For our calculation, it will be convenient to define
\begin{align}\label{Eq:qmnProp}
  \left \langle q_{\mu_1\nu_1}^{a_1}(x_1) q_{\mu_2 \nu_2}^{a_2}(x_2)
  \right\rangle = \delta_{a_1 a_2} D_{\mu_1\nu_1,\mu_2
    \nu_2}(x_1-x_2)
\end{align}
for $q_{\mu\nu}$ given in \eqref{Eq:qmn}. Using
\begin{align}
  q_{\mu\nu}(x) = a \left(\Delta_\mu q_\nu(x) - \Delta_\nu q_\mu(x)
  \right) 
\end{align}
we obtain
\begin{align}\label{Eq:qmnProp2}
  D_{\mu_1\nu_1,\mu_2 \nu_2} (z) = a^2\left( \delta_{\nu_1\nu_2}
    \Delta_{\mu_1} \bar \Delta_{\mu_2} +\delta_{\mu_1\mu_2}
    \Delta_{\nu_1} \bar \Delta_{\nu_2} -\delta_{\nu_1\mu_2}
    \Delta_{\mu_1} \bar \Delta_{\nu_2} -\delta_{\mu_1\nu_2}
    \Delta_{\nu_1} \bar \Delta_{\mu_2} \right) D(z)\,.
\end{align}
Using the background-field gauge ensures that $\Gamma$ is a
gauge-invariant functional of $A$. Assuming the background fields to
be small and slowly varying as usual, this implies that in the
continuum limit the lowest-order term is proportional to
$\tr F_{\mu\nu}^2$. In the following, we will only focus on terms of
order $A^2$ in $\Gamma$ (i.e., we do not explicitly check that the
linear term vanishes) and identify
$(\partial_\mu A_\nu - \partial_\nu A_\mu)^2$ with $F_{\mu\nu}^2$.

Note that the expectation value of a product of an odd number of $q$
fields vanishes. Therefore odd powers of $g$ are absent in
\eqref{Eq:gWgI-full}.

\subsubsection{Power counting}
\label{Sec:PowerCounting}

Counting powers of $\tgi$ and $\gamma$ in expectation values
(w.r.t.~the free action for the quantum field\footnote{The free action
  is obtained from the term of order $A^0q^2$ in $S_I^{(1,0)}$, which
  is of order $g^0\gamma^0$. Therefore, free propagators do not lead
  to additional factors of $g$ or $\gamma$, see
  \eqref{Eq:qPropagator} and~\eqref{Eq:Propagator}.}) of
products of the form
\begin{align}\label{Eq:PowerCountingExp}
  S_I^ {(l,2)} \prod_i \left(S_I^ {(n_{i},0)}\right)^{m_{i}}
  \,,\qquad S_I^{(l_1,1)} S_I^ {(l_2,1)} \prod_i \left(S_I^
    {(n_{i},0)}\right)^{m_{i}} \,,
\end{align}
we see that there are terms of order $A^2$ in $\Gamma[A]$ with
coefficients of order
\begin{align}\label{Eq:PowerCounting}
  \frac{1}{\tgi^2}\left(\frac{\tgi^2}{\gamma}\right)^n \left( 1
    +\mathcal O(\tgi^2) \right) = \frac 1{\gamma}
  \frac{1}{\Nb^{n-1}}\left( 1+ \mathcal O\left(\frac{\gamma}{\Nb}
    \right)\right), \qquad n\geq 0
\end{align}
with $n=l-1+\sum_i m_i(n_i-1)$ and $n=l_1+l_2-2+\sum_i m_i(n_i-1)$,
respectively.  We observe that only simple poles in $\gamma$ will
appear when we write the coefficients in terms of $\gamma$ and $\Nb$,
see section~\ref{Sec:Strategy}. Furthermore, the residue at the pole is
exclusively determined by expectation values of the
form~\eqref{Eq:PowerCountingExp}, where only the leading terms of
$S_I^{(n,k)}$ given in \eqref{Eq:SInk0} through \eqref{Eq:SInk2}
contribute. Subleading terms result in corrections of order $\gamma^m$
with $m\geq 0$ on the RHS of Eq.~\eqref{Eq:PowerCounting}.

Since the Wilson gauge action is given by the (implicitly defined) $n=1$
term in \eqref{Eq:S1gIm}, one might think that the leading terms of
$S_I^{(1,k)}$ with $k=0,1,2$ would contribute to the residue at the
pole in the effective action. However, the leading term of
$S_I^{(1,0)}$ is quadratic in the quantum field $q$ and therefore only
contributes to the free action. For $S_I^{(1,1)}$, the leading term is
linear in $q$ and therefore does not contribute to one-particle
irreducible diagrams (the same applies to the first subleading term of
$S_I^{(1,2)}$). Finally, the leading term of $S_I^{(1,2)}$ is just the
classical piece of the effective action.

It is obvious that diagrams containing measure vertices or ghost loops
cannot contribute to the coefficient of the pole in $\gamma$ in the
two-point function of the background field since these vertices
appear with powers of $\tgi^2=\gamma/\Nb$ without any accompanying
factors of $1/\gamma$.

Expectation values that involve only terms from $S_I^{(n=1)}$ and
$S_{\text{gf}}$ do not depend on $\gamma$ and are collected in
$c_m^{(W)}$ at order $(\tgi^2)^{m-1}$ on the RHS of
\eqref{Eq:gWgI-full}.

In summary, when the RHS of \eqref{Eq:gWgI-full} is written in
terms of $\Nb$ and $\gamma$, the coefficient of the pole in $\gamma$
is determined exclusively by expectation values of products of leading
terms of $S_I^{(n,k)}$ with $n\geq 2$. This means that we do not have
to consider mixing of the Wilson part of the induced action or
$S_{\text{gf}}$ with $n\geq 2$ terms in expectation values of
$ S_{\text{int}}^k $, see \eqref{Eq:S1gIm},
\eqref{Eq:GammaExpanded}, and~\eqref{Eq:PowerCountingExp}.

In order to determine the full two-loop coefficient $c_2(\gamma)$ in
the relation of $\gw$ and $\tgi$ in \eqref{Eq:gWgI} one also has
to take into account the renormalization of the gauge parameter, which
is obtained from the gluon self energy at one-loop order
\cite{Luscher:1995np}. However, we note that this does not result in a
contribution to the pole coefficient $c_{2,-2}$. The reason is again a
factor of $\tgi^2$ without any accompanying factor of
$1/\gamma$. The pole coefficients $c_{1,-1}$ and $c_{2,-2}$ can
therefore be determined exclusively from the two-point function of the
background field, by requiring $\Gamma_I[A]=\Gamma_W[A]$ at order
$A^2$ with $\Gamma[A]$ obtained through \eqref{Eq:GammaExpanded}.

\subsubsection{Effective action to one loop}\label{Sec:Gamma1loop}

To determine the coefficient of order $\tgi^0$ in
\eqref{Eq:gWgI-full}, we have to take into account terms of order
$A q^2$ and $A^2q^2$ from $S_I^{(n=1)}$ and $S_{\text{gf}}$. From
$S_I^{(n)}$ with $n\geq 2$, only the leading term of $S_I^{(2,2)}$
contributes, which is of order $A^2 q^2$. Therefore, terms from
$S_I^{(n=1)}$ and $S_{\text{gf}}$ determine $c_1(W)$ but do not
contribute to $c_1(\gamma)$ on the RHS of \eqref{Eq:gWgI-full}.
Hence, for the one-loop coefficient $c_1(\gamma)$, we only have to
calculate the expectation value of
\begin{align}
  S_I^{(2,2)} = - \frac{a^4}{\gamma} \sum_x \sum_{\mu,\nu}
  \Tr\Bigl[\left(A_{\mu\nu}(x)\right)^2
    \left(q_{\mu\nu}(x)\right)^2+\frac 12 A_{\mu\nu}(x) q_{\mu\nu}(x)
    A_{\mu\nu}(x) q_{\mu\nu}(x)\Bigr]\,,
\end{align}
see \eqref{Eq:SInk2}. Expanding in terms of $\SU(\Nc)$ generators,
we have\footnote{Sums over repeated color indices are always implied.}
\begin{align}
  \bigl\langle S_I^{(2,2)} \bigr\rangle & = - \frac{a^4}{\gamma}
  \sum_x \sum_{\mu,\nu} A^a_{\mu\nu}(x) A^b_{\mu\nu}(x) \bigl\langle
    q^c_{\mu\nu}(x) q^d_{\mu\nu}(x) \bigr\rangle \Tr \Bigl[ t_a t_b
    t_c t_d +\frac 12 t_a t_c t_b t_d \Bigr] \cr & = -
  \frac{a^4}{\gamma} \sum_x \sum_{\mu,\nu} A^a_{\mu\nu}(x)
  A^b_{\mu\nu}(x) D_{\mu\nu,\mu\nu}(0) \Tr \Bigl[ t_a t_b t_c t_c
    +\frac 12 t_a t_c t_b t_c \Bigr]
\end{align}
since the propagator for $q_\mu$ is diagonal in color space, see
\eqref{Eq:qmnProp}.  Making use of the identities provided in
appendix~\ref{Sec:ColorTraces} immediately results in
\begin{align}
  s^{(2,2)}_{ab} &= \Tr \Bigl[ t_a t_b t_c t_c +\frac 12 t_a t_c
    t_b t_c \Bigr] = \left(\frac{\Nc^2-1}{2 \Nc} - \frac 1{ 4
      \Nc}\right) \Tr(t_a t_b)
  = s^{(2,2)} \delta_{ab}\,,\\
  s^{(2,2)}&= \frac{2 \Nc^2-3}{8 \Nc}\,.
\end{align}
Due to $A_{\mu\mu} = 0$ we can restrict the sum over $\mu$ and $\nu$
to $\mu \neq \nu$. For this case, we obtain from
\eqref{Eq:qmnProp2}
\begin{align}
  D_{\mu\nu,\mu\nu}(0) = a^2 \left( \Delta_\mu \bar \Delta_\mu +
    \Delta_\nu \bar \Delta_\nu \right) D(0) \,.
\end{align}
Since $\Delta_\mu \bar \Delta_\mu D(0)$ is independent of $\mu$ due to
hypercubic symmetry, we get
\begin{align}\label{Eq:Dmnmn0}
  D_{\mu\nu,\mu\nu}(0) =a^2 \frac 2d \,\square D(0) = \frac{2}{d
    a^2}\,.
\end{align}
In the continuum limit, we can identify $a^{-1}A_{\mu\nu}(x)$ with the
field strength tensor $-F_{\mu\nu}(x)$ (since the effective action has
to be gauge invariant) and end up with
\begin{align}
  \bigl\langle S_I^{(2,2)} \bigr\rangle = - \frac 4d s^{(2,2)}
  \frac 1{\gamma} a^{4-d} \int d^dx \sum_{\mu,\nu} \Tr
  F_{\mu\nu}(x)^2+\ldots
\end{align}
Taking the continuum limit of $S_I^{(1,2)}$ in \eqref{Eq:SInk2} we
obtain
\begin{align}\label{Eq:Scl}
  S_{\text{cl}}[A] = \frac1{2 \tgi^2} a^{4-d} \int d^dx \sum_{\mu,\nu}
  \Tr F_{\mu\nu}(x)^2+\ldots \,,
\end{align}
which results in, see \eqref{Eq:gWgI-full},
\begin{align}\label{Eq:c1result}
  c_1(\gamma) = - \frac 8d s^{(2,2)} \frac 1{\gamma} = -
  \frac{2 \Nc^2-3}{\Nc d} \frac 1{\gamma}\,.
\end{align}
Using \eqref{Eq:c1} this means
\begin{align}\label{eq:c1-1}
  c_{1,-1}=-\frac{2 \Nc^2-3}{\Nc d}\,.
\end{align}

\subsubsection{Relevant two-loop contributions}\label{Sec:Gamma2loop}

From the general discussion in section~\ref{Sec:PowerCounting} we know
that, in order to obtain the coefficient $c_{2,-2}$ in \eqref{Eq:c2},
we only have to compute the contribution to
$\left\langle S_{\text{int}} -\frac 12 S_{\text{int}}^2 \right\rangle$
given by the leading term of
\begin{align}
  \bigl\langle S_I^{(3,2)} - S_I^{(2,0)}S_I^{(2,2)} -\frac12
    S_I^{(2,1)}S_I^{(2,1)} \bigr\rangle\,.
\end{align}

\paragraph{a) Expectation value of the $(3,2)$-term}

From \eqref{Eq:SInk2} we obtain
\begin{align}
  S_I^{(3,2)} = \frac{ a^{6} \tgi^2 }{\gamma^2} \sum_{x,\mu,\nu}\Tr
  \Bigl[& A_{\mu\nu}(x)^2 q_{\mu\nu}(x)^4 + A_{\mu\nu}(x)
  q_{\mu\nu}(x) A_{\mu\nu}(x) q_{\mu\nu}(x)^3 \cr &+ \frac 12
  A_{\mu\nu}(x) q_{\mu\nu}(x)^2 A_{\mu\nu}(x) q_{\mu\nu}(x)^2\Bigr]\,.
\end{align}
Since
\begin{align}
  \bigl\langle q^c_{\mu\nu}(x) q^d_{\mu\nu}(x) q^e_{\mu\nu}(x)
    q^f_{\mu\nu}(x) \bigr\rangle = \left(
    \delta_{cd}\delta_{ef}+\delta_{ce}\delta_{df}+\delta_{cf}\delta_{de}\right)
  D_{\mu\nu,\mu\nu}(0)^2\,,
\end{align}
see \eqref{Eq:qmnProp}, we need to calculate the trace
\begin{align}
  s^{(3,2)}_{ab}=\left(\delta_{cd}\delta_{ef}+\delta_{ce}\delta_{df}+
    \delta_{cf}\delta_{de}\right) \Tr\Bigl[t_a t_b t_c t_d t_e t_f +
    t_a t_c t_b t_d t_e t_f + \frac 12 t_a t_c t_d t_b t_e t_f\Bigr]
\end{align}
with sums over repeated color indices implied as usual. Using
\eqref{Eq:TrId1} and~\eqref{Eq:TrId2} we obtain
\begin{align}
  s^{(3,2)}_{ab}= s^{(3,2)} \delta_{ab}\,,\qquad
  s^{(3,2)} = \frac{5}{16}\left(\Nc^2-3+\frac
    3{\Nc^2}\right).
\end{align}
With \eqref{Eq:Dmnmn0} we then obtain
\begin{align}
  \bigl\langle S_I^{(3,2)} \bigr \rangle & =
  \frac{\tgi^2}{\gamma^2}\frac {8s^{(3,2)}}{d^2}
  a^2 \sum_{x,\mu,\nu} \Tr A_{\mu\nu}(x)^2\,,
\end{align}
and replacing $A_{\mu\nu}\to -a F_{\mu\nu}$ in the continuum limit
results in
\begin{align}\label{Eq:S32}
  \bigl\langle S_I^{(3,2)} \bigr \rangle = \frac{\tgi^2}{\gamma^2}
  \frac {8s^{(3,2)}}{d^2} a^{4-d} \int d^d x
  \sum_{\mu\nu} \Tr F_{\mu\nu}(x)^2+\ldots
\end{align}

\paragraph{b) Expectation value of the $(2,2;2,0)$-term}

For $\big\langle S_I^{(2,2)} S_I^{(2,0)} \big\rangle$ we have to
compute the 1PI connected expectation value
\begin{align}
  \bigl \langle q_{\mu\nu}^a(x) q_{\mu\nu}^b(x) q_{\mu\nu}^c(x)
    q_{\mu\nu}^d(x) q_{\rho\lambda}^e(y) q_{\rho\lambda}^f(y) \bigr
  \rangle_{\text{1PI-C}} = D_{\mu\nu,\rho\lambda}(x-y)^2
  D_{\mu\nu,\mu\nu}(0) f_{abcdef}
\end{align}
with
\begin{align}
  f_{abcdef} &= \delta_{ea}\left(\delta_{fb}\delta_{cd}+\delta_{fc}\delta_{bd}+\delta_{fd}\delta_{bc}\right)+
  \delta_{eb}\left(\delta_{fa}\delta_{cd}+\delta_{fc}\delta_{ad}+\delta_{fd}\delta_{ac}\right)\cr
  &\quad+
  \delta_{ec}\left(\delta_{fa}\delta_{bd}+\delta_{fb}\delta_{ad}+\delta_{fd}\delta_{ab}\right)+
  \delta_{ed}\left(\delta_{fa}\delta_{bc}+\delta_{fb}\delta_{ac}+\delta_{fc}\delta_{ab}\right).
\end{align}
We define the color factor as
\begin{align}
  s^{(2,2;2,0)}_{ij}= f_{abcdef} \Tr\left[t_a t_b t_c t_d\right]
  \Tr\Bigl[t_i t_j t_e t_f +\frac 12 t_i t_e t_j t_f\Bigr]
\end{align}
and obtain, by repeatedly using \eqref{Eq:TrId1}
and~\eqref{Eq:TrId2},
\begin{align}
  s^{(2,2;2,0)}_{ij} = s^{(2,2;2,0)} \delta_{ij} \,,\qquad
  s^{(2,2;2,0)} = 2\left(\frac{ 2\Nc^2-3}{4\Nc}\right)^2.
\end{align}
After some algebra we obtain
\begin{align}
  \sum_{\mu,\nu} \sum_{\rho,\lambda} \Tr A_{\rho
      \lambda}(y)^2\sum_x
  D_{\mu\nu,\rho\lambda}(x)^2 D_{\mu\nu,\mu\nu}(0) =
  \frac{8}{a^6 d^2} \sum_{\rho,\lambda} \Tr
    A_{\rho\lambda}(y)^2\,,
\end{align}
which leads to
\begin{align}
  \bigl\langle S_I^{(2,0)} S_I^{(2,2)} \bigr\rangle_{\text{1PI-C}}
  & = \frac{\tgi^2}{\gamma^2} \frac{4
      s^{(2,2;2,0)}}{d^2} a^2 \sum_{x,\mu,\nu} \Tr 
    A_{\mu\nu}(x)^2\,.
\end{align}
In the continuum limit this yields
\begin{align}\label{Eq:S2220}
  \bigl\langle S_I^{(2,0)} S_I^{(2,2)} \bigr\rangle_{\text{1PI-C}}
  =\frac{\tgi^2}{\gamma^2} \frac{ 4 s^{(2,2;2,0)}}{d^2}
  a^{4-d} \int d^d x \sum_{\mu,\nu} \Tr 
    F_{\mu\nu}(x)^2+\ldots
\end{align}

\paragraph{c) Expectation value of the $(2,1;2,1)$-term}

For the contribution of
$\big\langle S_I^{(2,1)} S_I^{(2,1)} \big\rangle$ we only need to
compute the 1PI connected part
\begin{align}
  \bigl \langle q_{\mu\nu}^a(x) q_{\mu\nu}^b(x) q_{\mu\nu}^c(x)
    q_{\rho\lambda}^d(y) q_{\rho\lambda}^e(y) q_{\rho\lambda}^f(y)
  \bigr\rangle_{\text{1PI-C}} =
  D_{\mu\nu,\rho\lambda}(x-y)^3g_{abcdef}
\end{align}
with
\begin{align}
  g_{abcdef}= \delta_{ad}\delta_{be}\delta_{cf} +
  \delta_{ad}\delta_{bf}\delta_{ce} +
  \delta_{ae}\delta_{bd}\delta_{cf} +
  \delta_{ae}\delta_{bf}\delta_{cd} +
  \delta_{af}\delta_{bd}\delta_{ce} +
  \delta_{af}\delta_{be}\delta_{cd} \,.
\end{align}
Thus, we need to evaluate
\begin{align}
  s_{ij}^{(2,1;2,1)}= g_{abcdef}\Tr\left[t_i t_a t_b t_c\right]
  \Tr\left[t_j t_d t_e t_f\right].
\end{align}
Using \eqref{Eq:TrId1} and~\eqref{Eq:TrId2} we obtain after some
algebra
\begin{align}
  s^{(2,1;2,1)}_{ij} = s^{(2,1;2,1)} \delta_{ij} \,,\qquad
  s^{(2,1;2,1)} = \frac{\Nc^4-6\Nc^2+18}{16 \Nc^2}\,.
\end{align}
We assume the background field to be slowly varying
\cite{Dashen:1980vm} and expand $A_{\rho\lambda}(y)A_{\mu\nu}(x)$
around a common point, e.g., $x$, effectively substituting
$A_{\rho\lambda}(y)\to A_{\rho\lambda}(x)$ at leading order. Then we
only need to compute
\begin{align}
  \sum_{\mu\neq \nu,\rho\neq\lambda}\Tr\left[A_{\mu\nu}(x)
    A_{\rho\lambda}(x)\right] \sum_y
  D_{\mu\nu,\rho\lambda}(y)^3\,.
\end{align}
Since the effective background-field action is gauge invariant, all
contributions with $\{\mu,\nu\}\neq\{\rho,\lambda\}$ vanish.  After
some algebra we obtain
\begin{align}
  \sum_{\mu\neq \nu,\rho\neq\lambda}\Tr\left[A_{\mu\nu}(x)
    A_{\rho\lambda}(x)\right] \sum_y
  D_{\mu\nu,\rho\lambda}(y)^3 = a^{-6} C_d
  \sum_{\mu\neq\nu} \Tr A_{\mu\nu}(x)^2
\end{align}
with
\begin{align}
  C_d& = \frac{4}{d-1}\left(\frac{3}{d^2} -4 (4-d) J_d\right), \\
  J_d& = \frac 1 8 \int_{-\pi}^\pi \frac{d^dk}{(2\pi)^d}
  \frac{d^dq}{(2\pi)^d} \frac{
    \left(\sin(k_1)+\sin(q_1)-\sin(k_1+q_1)\right)^2 } {
    \sum_{\gamma,\mu,\rho}(1-\cos(k_\gamma)) (1-\cos(q_\mu))
    (1-\cos(k_\rho+q_\rho)) }\,.\label{Eq:Jd}
\end{align}
For $d=2$, we find after some algebra $J_{d=2}=\frac1{32}$. For $d=3$
we have evaluated the integral $J_d$ numerically and obtained
$J_{d=3}\approx 0.0085535415$. This results in
\begin{align}
  C_{d=4}=\frac 14\,,\qquad C_{d=3}\approx
  0.59823833\,,\qquad C_{d=2}=2
\end{align}
and leads to
\begin{align}\label{Eq:S2121}
  \frac 12 \bigl\langle S_I^{(n=2,k=1)} S_I^{(n=2,k=1)}
  \bigr\rangle_{\text{1PI}} &= \frac{\tgi^2}{\gamma^2}
  s^{(2,1;2,1)} C_d a^2 \sum_{x,\mu,\nu} \Tr 
    A_{\mu\nu}(x)^2 \notag\\ & =
  \frac{\tgi^2}{\gamma^2} s^{(2,1;2,1)} C_d a^{4-d}
  \int d^dx \sum_{\mu,\nu} \Tr F_{\mu\nu}(x)^2+\ldots
\end{align}

\paragraph{d) Coefficient $c_{2,-2}$}

From \eqref{Eq:c2}, \eqref{Eq:gWgI-full}, \eqref{Eq:Scl},
\eqref{Eq:S32}, \eqref{Eq:S2220}, and~\eqref{Eq:S2121} we finally
obtain
\begin{align}
  c_{2,-2}&=\frac 8{d^2}\bigl(2 s^{(3,2)} - s^{(2,2;2,0)}\bigr)-
  2 s^{(2,1;2,1)} C_d\cr &= \frac{\Nc^4-3\Nc^2+6}{d^2 \Nc^2} -
  \frac{\Nc^4-6 \Nc^2+18}{2 \Nc^2 (d-1)} \left(\frac 3{d^2}-4(4-d)
    J_d\right)
\end{align}
and in particular
\begin{align}
  c_{2,-2}|_{d=4} &= \frac{\Nc^4-6}{32 \Nc^2}\,,\\
  c_{2,-2}|_{d=3} &= \frac{\Nc^4+6 \Nc^2-30}{36 \Nc^2}+\frac{\Nc^4-6\Nc^2+18}{\Nc^2} J_3\,,\label{eq:c2-2d3}\\
  c_{2,-2}|_{d=2} &= \frac34-\frac3{\Nc^2}
\end{align}
with $J_d$ from \eqref{Eq:Jd} and $J_{3}\approx 0.0085535415$.

For $d=2$, the results for $c_{1,-1}$ and $c_{2,-2}$ are consistent
with \eqref{eq:alpha-gw-2d}, which was obtained by matching the
character expansions of the plaquette weight functions for the Wilson
action and the induced action in the continuum limit.

\subsection{Comparison with numerical results}

\begin{figure}
  \begin{center}
    \includegraphics{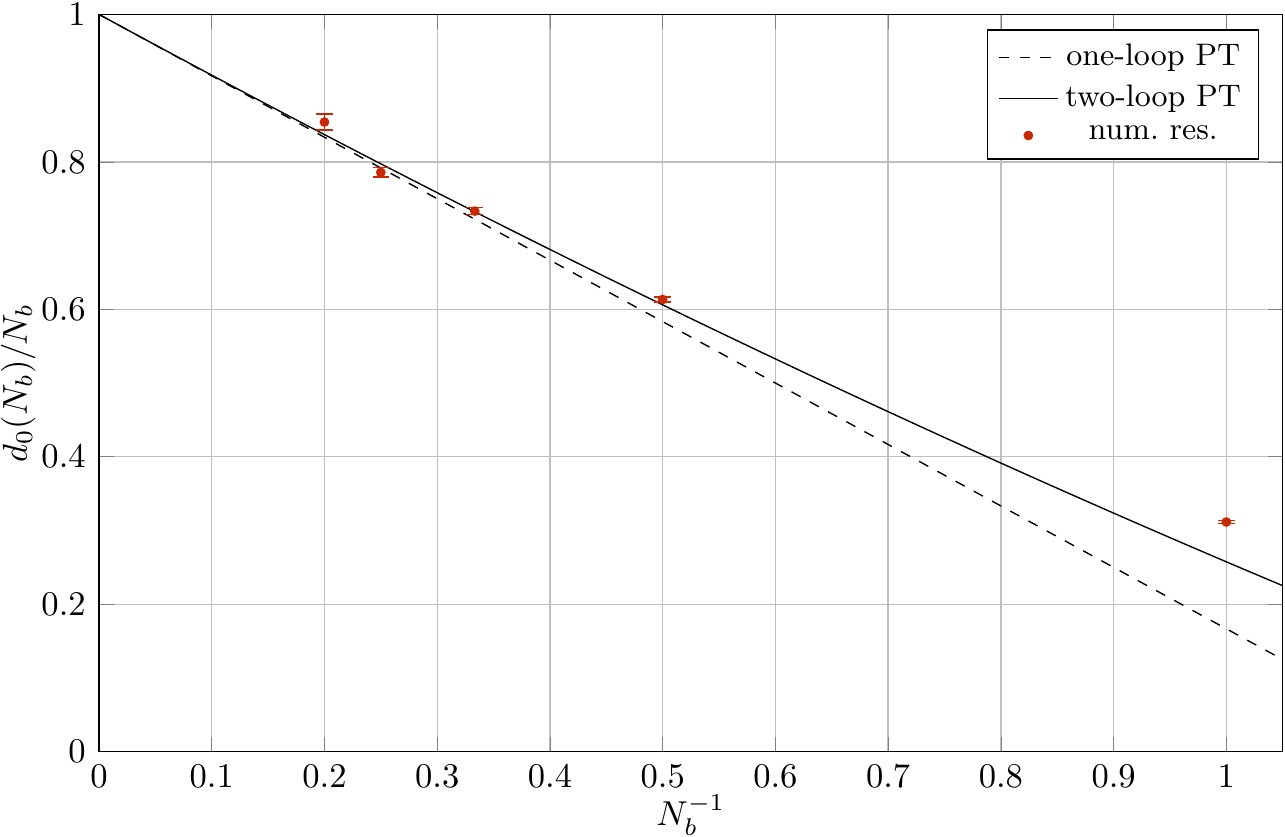}
    \caption{Perturbative and numerical results for $d_0/\Nb$ in $d=3$
      with $\Nc=2$.}
    \label{Fig:comparison}
  \end{center}
\end{figure}

Using the methods and results introduced in \cite{Brandt:2014rca}, we
determined the coefficient $d_0$ in \eqref{Eq:gItilde} numerically
through simulations with both Wilson and induced gauge action for
$\Nc=2$ in three dimensions.\footnote{This was done by first matching
  the bare couplings of both approaches through the determination of
  the Sommer parameter $r_0$ \cite{Sommer:1993ce}. Then the data were
  fitted to \eqref{eq:match}, including the $\mathcal O(\gamma)$
  term. We have simulated at couplings corresponding to
  $0.116\le\gamma\le3.26$. The details of the simulations and the
  numerical results will be discussed in \cite{BLW:II}.}  Using
\eqref{eq:c1-1} and \eqref{eq:c2-2d3} for $d=3$ and $\Nc=2$, the
perturbative expansion of $d_0$ in \eqref{Eq:d0pert} reads
\begin{align}\label{eq:d032}
  \frac{d_0(\Nb)}{\Nb} =1 - \frac{5}{6 \Nb} +
  \frac{0.0908283}{\Nb^2}+\mathcal O(\Nb^{-3})\,.
\end{align}
The numerical results are shown in figure~\ref{Fig:comparison}
together with the perturbative results. Note that the latter were
derived assuming large $\Nb$ and $\gamma\ge4$. Nevertheless we observe
very good agreement even for small $\Nb$ and small $\gamma$, i.e.,
outside the domain of validity of \eqref{eq:d032}.

Note also that the numerical results include the value $\Nb=1$, which
is outside the bound \eqref{eq:cont} (for the continuum limit to be
equivalent to YM theory in $d=2$) but inside the bound
\eqref{eq:delta} (for the continuum limit to exist at all). The fact
that the corresponding data point is close to the perturbative results
is consistent with our expectation (formulated in
section~\ref{Sec:cont3d4d}) that the continuum limit of the induced
theory is equivalent to YM theory in $d>2$ whenever the continuum
limit actually exists.

\section{Conclusions}

In this paper we have explored the novel approach of Budzcies and
Zirnbauer \cite{Budczies:2003za} to induced QCD, which represents a
major step forward compared to earlier approaches since it requires
only a small number $\Nb$ of auxiliary bosons. We slightly
reformulated the action to solve a trivial sign problem present in the
original formulation. We then extended the analysis of
\cite{Budczies:2003za} from gauge group $\U(\Nc)$ to $\SU(\Nc)$.  The
latter case is of particular interest since it includes the gauge
group of QCD. We derived refined bounds on $\Nb$, admitting also
non-integer values, for the induced theory to have a continuum limit
at fixed $\Nb$ and for this continuum limit to be in the universality
class of YM theory in two dimensions. We conjectured that in higher
dimensions the latter bounds can be relaxed. We also performed a
perturbative calculation using the background-field technique to match
the bare coupling of the induced theory to the standard lattice
coupling. Formally, the result of this calculation is only valid in
the continuum limit $\Nb\to\infty$ at fixed $\alpha\le\frac13$. The
latter condition excludes the ``interesting'' continuum limit
$\alpha\to1$ at fixed $\Nb$. However, by comparing to data from
lattice-gauge simulations near the continuum limit, we observe very
good agreement also for parameters outside the formal range of
validity. This leads us to conjecture that the perturbative results
are also valid in the ``interesting'' continuum limit $\alpha\to1$.

In future work, we will present detailed numerical evidence from
lattice simulations in three and four dimensions that standard lattice
gauge theory and induced theory (at fixed $\Nb$) have the same
continuum limit, and that away from the continuum limit they differ
only by relatively small lattice artifacts. The numerical simulations
include quantities at both zero and non-zero temperature.

Having presented analytical and numerical evidence in support of the
induced theory, an important question is to what extent this new
approach is useful in the sense that it leads to better simulation
algorithms or new formulations that would allow us to go beyond what
is possible in the standard formulation. To this end we will explore a
dual formulation of lattice gauge theory, including fermions, in which
the gauge field is integrated out first. This can be done since it
only appears linearly in the action. After integrating out the
fermions, the remaining path integral over the auxiliary boson fields
involves only gauge-invariant objects. It will be interesting to see
whether a worm-like algorithm can be constructed for this dual
formulation.

\appendix

\section{Color traces}\label{Sec:ColorTraces}
The traceless generators $t_a$ in the fundamental representation of
$\SU(\Nc)$ are normalized according to
\begin{align}\label{Eq:GenNorm}
  \Tr(t_a t_b) = \frac 12 \delta_{ab}
\end{align}
and obey
\begin{align}
  \sum_c (t_c)_{ij} (t_c)_{kl} = \frac 12 \Big(
    \delta_{il}\delta_{jk} - \frac 1 {\Nc}
    \delta_{ij}\delta_{kl}\Big)\,,\qquad\sum_c t_c t_c =
  \frac{\Nc^2-1}{2 \Nc}\,.
\end{align}
For arbitrary matrices $A$ and $B$ we therefore have
\begin{align}
  \sum_c \Tr\left( A t_c B t_c\right)&=\frac 12 \left(\Tr
    A\right)\left(\Tr B\right)
  -\frac 1{2\Nc} \Tr\left(AB\right), \label{Eq:TrId1}\\
  \sum_c \Tr\left(A t_c\right) \Tr \left(B t_c\right) &= -\frac
  1{2\Nc} \left(\Tr A\right)\left(\Tr B\right) +\frac 12
  \Tr\left(AB\right)\,.\label{Eq:TrId2}
\end{align}

\section{\boldmath Character expansion for $\SU(2)$ and $\U(1)$}

\subsection[$\SU(2)$]{\boldmath $\SU(2)$}
\label{App:su2}

Every element $U$ of $\SU(2)$ can be diagonalized according to
$U=V\diag(e^{i\phi},e^{-i\phi})V^\dagger$ with $V$ unitary and
$\phi\in[-\pi,\pi]$. In the character expansion we only integrate over
class functions, which are independent of $V$. It therefore suffices
to integrate over $U=e^{i\phi \sigma_3}$ with measure
\begin{align}
  d\mu(\phi)=\frac{1}{\pi} \sin^2\phi\,.
\end{align}
The dimension $d_k$, the quadratic $\SU(2)$ Casimir operator $C_2(k)$,
and the character $\chi_k$ for the irreducible representation
corresponding to a one-row Young diagram with $k$ boxes are given by
\begin{align}
  d_k&=k+1\,,\\
  C_2(k) &= \frac 14 k (k+2) \,,\\
  \chi_k(\phi)&=\frac{\sin((k+1)\phi)}{\sin \phi}\,.
\end{align}
Since
\begin{align}
  \det\left(1-\frac \alpha 2(U+U^\dagger)\right) = (1- \alpha
  \cos\phi)^2 \propto \left( (1-b e^{i\phi})(1-b
    e^{-i\phi})\right)^2
\end{align}
with
\begin{align}
  \alpha=\frac{2b}{1+b^2}\,,\qquad b = \frac 1\alpha
  \left(1-\sqrt{1-\alpha^2}\right),
\end{align}
we write the unnormalized weight function as
\begin{align}
  \bar \omega(\phi)&= \frac1{\left( \left(1-b
        e^{i\phi}\right)\left(1-b e^{-i\phi}\right)\right)^{2
      \Nb}}
\end{align}
for convenience. Expanding in characters, we have
\begin{align}
  \bar \omega(\phi)=\sum_{n=0}^{\infty} \bar c_n \chi_n(\phi) \,,
  \qquad \bar c_n = \int_{-\pi}^{\pi} d\mu(\phi)\, \bar\omega(\phi)
  \chi_n(\phi)\,.
\end{align}
For the properly normalized weight function
\begin{align}
  \omega (\phi) = \frac {\bar \omega(\phi)}{\int d\mu(\phi)\, \bar
    \omega(\phi) }= \sum_{n=0}^{\infty} c_n \chi_n(\phi)
\end{align}
we obtain the expansion coefficients $c_n=\bar c_n / \bar c_0$.
Using the series expansion
\begin{align}
  \frac1{\left(1-b e^{i\phi}\right)^{2 \Nb}}=\sum_{k=0}^\infty {2
    \Nb+k-1 \choose k} b^k e^{i k \phi} \,,\qquad {n\choose
    m}=\frac{\Gamma(n+1)}{\Gamma(m+1)\Gamma(n-m+1)}\,,
\end{align}
which is valid for $b<1$, i.e., $\alpha <1$, and in which $2 \Nb$ is
not necessarily restricted to integer values, we obtain after some
algebra
\begin{align}\label{Eq:cn}
  \bar c_n = (n+1) b^n \sum_{m=0}^\infty b^{2m} \frac1{m+n+1}
  {2 \Nb-2 + m \choose m }{2 \Nb -1 + m +n \choose m +n }\,.
\end{align}
If $2 \Nb\in \mathbbm{N}$, the infinite sum can be calculated
analytically. In the following we consider the first few cases
explicitly.

(i) For $\Nb=\frac12$ we obtain
\begin{align}
  \bar c_n=b^n\,,\qquad c_n=b^n\,.
\end{align}
In the limit $\alpha \to 1$ (i.e., $b \to 1$) we have $c_n=1\neq d_n$,
and therefore $\omega$ does not reduce to a $\delta$-function on the
$\SU(2)$ manifold, as expected for $\Nb <\frac34$.

(ii) For $\Nb=1$ we obtain
\begin{align}
  \bar c_n&=(n+1)\frac{b^n}{1-b^2}\,,\\
  c_n&=(n+1)b^n=d_n(1-n(1-b)+\ldots)=d_n(1-n
       \sqrt{2(1-\alpha)}+\ldots)\,.
\end{align}
Taking $\alpha \to 1$, the leading term is given by $d_n$, but the
coefficient of the next-order term is not proportional to $C_2$, as
expected for $\frac 34 \leq \Nb < \frac 54$.

(iii) For $\Nb=\frac 32$ we obtain
\begin{align}
  \bar c_n&=(n+1) \frac{b^n}{(1-b^2)^3}\left(1+\frac n2(1-b^2)\right)\,,\\
  c_n&=d_n(1- 2 C_2(n)(1-b)^2+\ldots)=d_n(1- 4 C_2(n)
       (1-\alpha)+\ldots)
\end{align}
in agreement with \eqref{eq:SU-NLO} and~\eqref{Eq:TrHsqSU2}.

(iv) For $\Nb=2$ we obtain
\begin{align}
  \bar
  c_n&=\frac{(n+1)b^n}{6\left(1-b^2\right)^5}\left((n+3)(n+2)+2(n+3)(1-n)b^2
       -n(1-n)b^4\right),\\
  c_n &=d_n\left(1-\frac 23 C_2(n) (1-b)^2+\ldots\right)
        =d_n\left(1-\frac 43 C_2(n) (1-\alpha)+\ldots\right)
\end{align}
in agreement with \eqref{eq:SU-NLO} and~\eqref{Eq:TrHsqSU2}.

For non-integer $2\Nb$, $c_n$ can be easily expanded around $\alpha=1$
numerically. Our numerical results are consistent with
\begin{align}
  \frac{c_n}{d_n}=
  \begin{cases}
    1-\frac{4}{4 \Nb -5} C_2(n) (1-\alpha)+
    \mathcal O\left((1-\alpha)^{\min(\frac 32,\,2\Nb-\frac 32)}\right)  & \text{for }  \Nb>\frac 54\,,\\
    1+2 C_2(n)(1-\alpha)\log(1-\alpha)+\ldots & \text{for } \Nb=\frac 54\,,\\
    1- f(n)(1-\alpha)^{2 \Nb-\frac 32}+ \ldots  & \text{for }  \frac 34<\Nb<\frac 54\,,\\
    b^n
    \frac{\Gamma(\frac32+n)\,{}_2F_1(\frac12,\frac32+n;2+n;b^2)}{\Gamma(n+2)\Gamma(\frac
      32)\, {}_2F_1(\frac12,\frac32;2;b^2)}
    =1-\frac{g(n)}{\log(1-\alpha)}+\ldots & \text{for } \Nb=\frac 34\,,
  \end{cases}
\end{align}
where $f(n)$ and $g(n)$ are not proportional to $C_2(n)$.

\subsection[$\U(1)$]{\boldmath $\U(1)$}
\label{App:u1}

The irreducible representations of $\U(1)$ are one-dimensional and
characterized by $\chi_n(\phi)=e^{i n \phi}$ with $n\in \mathbbm Z$.
Proceeding analogously to the $\SU(2)$ case discussed in detail in the
previous subsection, we obtain
\begin{align}
\bar c_n=b^{|n|} \sum_{m=0}^\infty b^{2m}{\Nb+m-1 \choose m}{\Nb+m+|n|-1 \choose m+|n|}\,.
\end{align}
For $\Nb \in \mathbbm N$, the infinite sum can be calculated
analytically, and we obtain for the first few cases
\begin{align}
c_n=\frac{\bar c_n}{\bar c_0}=
\begin{cases}
b^{|n|}=1-|n| \sqrt \gamma + \ldots&\text{for }\Nb=1\,,\\
1-\frac \gamma 2 n^2+\ldots&\text{for }\Nb=2\,,\\
1-\frac \gamma 6 n^2+\ldots&\text{for }\Nb=3\,,\\
1-\frac \gamma {10} n^2+\ldots&\text{for }\Nb=4\,.
\end{cases}
\end{align}

\section{Explicit expansion of the induced gauge action}\label{App:ExplicitExpansion}
For an unoriented plaquette $p=(x,\mu<\nu)$, we write
\begin{align}\label{Eq:Wmn}
  U_p= U_{\mu\nu}(x)= 
  U_\nu^\dagger(x)U_\mu^\dagger(x+\nu)U_\nu(x+\mu)U_\mu(x)\,.
\end{align}
Using \eqref{Eq:UqA}, we then expand in powers of the background
field (up to quadratic order),
\begin{align}\label{Eq:DefC012} U_{\mu\nu}(x)+U_{\mu\nu}(x)^\dagger-2=C^{(0)}_{\mu\nu}(x)+C^{(1)}_{\mu\nu}(x)+C^{(2)}_{\mu\nu}(x)
  +\mathcal O \left(A^3\right),
\end{align}
where the $C^{(i)}_{\mu\nu}(x)$ are of order $A^i$.  Due to the
invariance of the trace under cyclic permutations, we then get for
$n\geq 2$ (omitting obvious indices and arguments)
\begin{align}\label{Eq:WexpansionC}
  \Tr\, \bigl(U_p+U_p^\dagger-2\bigr)^n=\Tr\, \biggl[ &
  \big(C^{(0)}\big)^n + n C^{(1)} \big(C^{(0)}\big)^{n-1} + n
  C^{(2)} \big(C^{(0)}\big)^{n-1} \cr & +\frac n2
  \sum_{m=0}^{n-2} C^{(1)} \big(C^{(0)}\big)^m C^{(1)}
  \big(C^{(0)}\big)^{n-m-2}\biggr]+\mathcal O\left(A^3\right).
\end{align}
The last term can also be rewritten as
\begin{align}\label{Eq:Trcomb}
  \frac n2 \Tr \sum_{m=0}^{n-2} C^{(1)} \big(C^{(0)}\big)^m C^{(1)}
  \big(C^{(0)}\big)^{n-m-2} &= - \frac n 2
  \delta_{0,n\,\text{mod}\,2} \Tr C^{(1)} \big(C^{(0)}\big)^{\frac
    n 2-1} C^{(1)} \big(C^{(0)}\big)^{\frac n 2 -1}\cr &\quad +n
  \Tr \sum_{m=0}^{\lfloor \frac n 2 \rfloor-1} C^{(1)}
  \big(C^{(0)}\big)^m C^{(1)} \big(C^{(0)}\big)^{n-m-2}.
\end{align}
Next, we expand the $C^{(i)}$ to leading order in $gq$.  For arbitrary
non-commuting matrices $M_i$ ($i=1,\ldots, k$) we have
\begin{align}
  e^{M_1}e^{M_2}\cdots e^{M_k} + e^{-M_k}\cdots e^{-M_2}e^{-M_1}-2 & =
  \sum_{i=1}^k M_i^2+ \sum_{i<j} M_i M_j + \sum_{i>j} M_i M_j +
  \mathcal O(M^3)\notag\\ 
  & = \biggl(\sum_{i=1}^k M_i \biggr)^2 +\mathcal O(M^3)\,. \label{Eq:M}
\end{align}
If we apply \eqref{Eq:M} to the LHS of \eqref{Eq:DefC012} we obtain
\begin{align}\label{eq:Ulhs}
  U_{\mu\nu}(x)+U_{\mu\nu}(x)^\dagger-2 = \left[(iag) q_{\mu\nu}(x) +
    (ia) A_{\mu\nu}(x)\right]^2 + \mathcal O(A^3,gq A^2,(gq)^2
  A,(gq)^3)
\end{align}
with $q_{\mu\nu}(x)$ and $A_{\mu\nu}(x)$ defined in \eqref{Eq:qmn}
and~\eqref{Eq:Amn}, respectively. The leading-order terms in $C^{(i)}$
can be read off from \eqref{eq:Ulhs} as
\begin{align}
  C_{\mu\nu}^{(0)}(x) & = -a^2g^2q_{\mu\nu}(x)^2+\mathcal O((gq)^3)\,,\\
  C_{\mu\nu}^{(1)}(x) & =-a^2 g \left( q_{\mu\nu}(x) A_{\mu\nu}(x) +
    A_{\mu\nu}(x) q_{\mu\nu}(x) \right)
  + \mathcal O((gq)^2)\,,\\
  C_{\mu\nu}^{(2)}(x) &=-a^2A_{\mu\nu}(x)^2+\mathcal
  O(gq)\,.
\end{align}
Analogously, the leading terms in the expansion~\eqref{Eq:WexpansionC}
can be read off from the expansion of
$\Tr(iag q_{\mu\nu} + iaA_{\mu\nu})^{2n}$ up
to $\mathcal O(A^2)$. Explicitly, we have
\begin{align}
  \Tr\, \bigl(C^{(0)}_{\mu\nu}(x)\bigr)^n &= (-1)^n
  a^{2n}g^{2n} \Tr q_{\mu\nu}(x)^{2n}+
  \mathcal O\left((gq)^{2n+1}\right),
  \label{Eq:TrC0leading}
  \\
  n \Tr\, \Bigl[ C^{(1)}_{\mu\nu}(x)
    \bigl(C^{(0)}_{\mu\nu}(x)\bigr)^{n-1} \Bigr] & = (-1)^n a^{2n}
  g^{2n-1} 2n \Tr\,\Bigl[ A_{\mu\nu}(x)
    q_{\mu\nu}(x)^{2n-1} \Bigr]+ \mathcal
  O\left((gq)^{2n}\right),
  \label{Eq:TrC1leading}
\end{align}
and, combining terms of order $A^2$,
\begin{align}
  & \Tr\biggl[n C^{(2)}_{\mu\nu}(x)
    \bigl(C^{(0)}_{\mu\nu}(x)\bigr)^{n-1} +\frac n2 \sum_{m=0}^{n-2}
    C^{(1)}_{\mu\nu}(x) \bigl(C^{(0)}_{\mu\nu}(x)\bigr)^m
    C^{(1)}_{\mu\nu}(x)
    \bigl(C^{(0)}_{\mu\nu}(x)\bigr)^{n-m-2}\biggr]\cr &=(-1)^n a^{2n}
  g^{2n-2} n\sum_{m=0}^{2n-2} \Tr \left[ A_{\mu\nu}(x)
    q_{\mu\nu}(x)^m A_{\mu\nu}(x)
    q_{\mu\nu}(x)^{2 n-m-2}\right]+ \mathcal
  O\left((gq)^{2n-1}\right) \cr & = (-1)^n a^{2n} g^{2n-2} 2n \Tr
  \Biggl[\sum_{m=0}^{n-1} A_{\mu\nu}(x) q_{\mu\nu}(x)^m
  A_{\mu\nu}(x) q_{\mu\nu}(x)^{2 n-m-2} 
  -\frac12 \left(A_{\mu\nu}(x)q_{\mu\nu}(x)^{n-1}\right)^2 \Biggr] \cr
  &\quad + \mathcal O\left((gq)^{2n-1}\right),
  \label{Eq:TrC2leading}
\end{align}
where we have made use of \eqref{Eq:Trcomb} to obtain the last
expression.  Taking into account the prefactors in
\eqref{Eq:S1gIm}, we end up with $S_I^{(n,k=0,1,2)}$ given in
\eqref{Eq:SInk0} through \eqref{Eq:SInk2}.

\acknowledgments

This work was supported by DFG in the framework of SFB/TRR-55.  We
thank Christoph Lehner for useful discussions.  We also thank Philippe
de Forcrand and H\'elvio Vairinhos for discussions on their
unpublished work.

\bibliographystyle{JHEP}
\bibliography{spires}

\end{document}